\documentclass[aps, pra, twocolumn, showpacs,nofootinbib,superscriptaddress,longbibliography]{revtex4-1}

\usepackage{psfrag,graphicx}
\usepackage{dcolumn}
\usepackage{bm}
\usepackage{amsfonts,amssymb,amsmath,amsthm}
\usepackage{lipsum}
\usepackage{dsfont}
\usepackage{physics}
\usepackage{comment}
\usepackage[justification=raggedright,font=small,labelfont=bf,format=plain]{caption}
\usepackage{subcaption}
\usepackage[usenames, dvipsnames]{color}
\usepackage{xcolor}
\usepackage{comment}
\usepackage{mathtools}
\usepackage{todonotes}
\usepackage{tikz,siunitx}
\usetikzlibrary{shapes.geometric,shapes.symbols}

\newtheorem{theorem}{Theorem}[section]

\newtheorem{lemma}[theorem]{Lemma}

\theoremstyle{definition}

\newcommand{\Imperial}{Blackett Laboratory, Imperial College London, SW7 2AZ, United Kingdom}
\newcommand{\Sydney}{Center for Engineered Quantum Systems, Macquarie University, NSW 2109, Australia}
\newcommand{\Dresden}{Helmholtz-Zentrum Dresden-Rossendorf, Bautzner Landstraße 400, 01328 Dresden, Germany}

\newcommand{\old}[1]{{\color[rgb]{0,0.8,0}{..}}}

\newcommand{\be}{\begin{equation}}
\newcommand{\ee}{\end{equation}}
\newcommand{\ovl}[2]{\langle #1 | #2 \rangle}

\begin{document}

\title{Scalable quantum control and non-abelian anyon creation in the Kitaev honeycomb model}
\date{2022}

\author{Omar Raii}\affiliation{\Imperial}\affiliation{\Sydney}
\author{Florian Mintert}\affiliation{\Imperial}\affiliation{\Dresden}
\author{Daniel Burgarth}\affiliation{\Sydney}

\begin{abstract}
The Kitaev honeycomb model is a system allowing for experimentally realisable quantum computation with topological
protection of quantum information. Practical implementation of quantum information processing
typically relies on adiabatic, \textit{i.e.} slow dynamics. Here we show that the restriction to adiabatic
dynamics can be overcome with optimal control theory, enabled by an extension of the fermionization
of the Kitaev honeycomb model to the time-dependent case. Moreover we present a quantum control method that is applicable to large lattice models due to sub-exponential scaling.

\end{abstract}

\maketitle

\section{Introduction}

Non-abelian anyons are the foundational theoretical tools for topological quantum computation. These generalizations of bosons and fermions, which can only exist in two-dimensional systems \cite{Kitaev2006, Wilczek1982QuantumParticles}, allow for a topological form of gate implementation due to their non-trivial braiding statistics \cite{Sarma2006,Pachos2012,Kitaev1997}.
Logical qubits are encoded non-locally and anyons may be braided around one another and subsequently fused to carry out computations and measurements respectively as part of the overall implementation of a quantum algorithm \cite{Pachos2012,Brennen2007WhyAnyons}. 

The Kitaev honeycomb model is a notable example of a relatively simply defined system with non-trivial topological order \cite{Pachos2012, Kitaev1997}. The simplicity of the model's definition has led to multiple proposals in recent years for experimental realization \cite{Knolle2019ALiquids,Takagi2019ConceptLiquids}. In this system, anyons manifest as  vortices introduced into the model which may be fused to create fermionic excitations corresponding to anyonic fusion rules known as Ising anyons \cite{Lahtinen2011InteractingModel,GenetayJohansen2021FibonacciModels}. Anyonic braiding itself and indeed any form of particles being interchanged is generally assumed to be an adiabatic process \cite{Cheng2011NonadiabaticSuperconductors} so that unwanted excitations may be safely suppressed. 

In practice, however, the restriction to adiabatic dynamics is typically conflicting with the requirement to realize all operations on a time-scale that is short compared to the system's coherence time~\cite{Lai2020DecoherenceOperations}.
Quantum control has proven successful in speeding up adiabatic evolution in a wide range of scenarios \cite{He2016EfficientDriving,Zhou2017AcceleratedSystem,Xu2019BreakingGeodesics,Wang2012AdiabaticControl,Simsek2021QuantumInvariant} suggesting its suitability for anyon creation in topological systems.
Common optimal control techniques, however, are limited in their applicability to the Kitaev honeycomb model.
Due to the exponential scaling of composite quantum systems, numerical simulations of the time-dependent Kitaev honeycomb model are only possible for very small system sizes. fermionization of the Kitaev honeycomb model allows for improved scaling and solving larger systems but thus far this has been restricted to systems with time-independent Hamiltonians.
The scope of this article is to use quantum control in applying fermionization within the context of a time-dependent version of the model. This will demonstrate that optimal control does indeed provide access to faster-than-adiabatic anyon creation.

Sec. II of this paper provides a brief overview of anyons in the Kitaev honeycomb model to setup the operators and terminology required for the control problem. Sec. III describes quantum control and pulse optimization for anyon creation and sets up the key result on fermionization, which is proven in the appendix. Sec. IV presents explicit numerical results of the optimal control problem defined. An overall summary of results and conclusions are presented in the final section.

\section{Anyons in the Kitaev honeycomb model}

Numerous methods have been demonstrated which allow for solving the Kitaev honeycomb model including that of Feng et al. \cite{Feng2007TopologicalModel} using the Jordan-Wigner transformation as well as that of Kells et al. \cite{Kells2009DescriptionStabilizers,Brennan2018TheGgeq2} which employs a more specific Jordan-Wigner type mapping that describes the system in terms of hard-core bosons which are then fermionized. Such methods have many advantages and have applications to other lattice structures \cite{Kells2010ExactLiquid}. The method most suitable for our work in extending to a time-dependent model is Kitaev's original Majorana fermionization procedure and in this section we review this method \cite{Kitaev2006,Knolle2016DynamicsLiquid}. This is followed by a demonstration of how vortex creation is implemented within the model \cite{Pachos2012,Lahtinen2017AComputation} and how this corresponds to creation of non-abelian anyons.

\subsection{Diagonalising the honeycomb}
\label{Majoranadiag}
\begin{figure}[h]
\centering
    \includegraphics[width=0.8\linewidth]{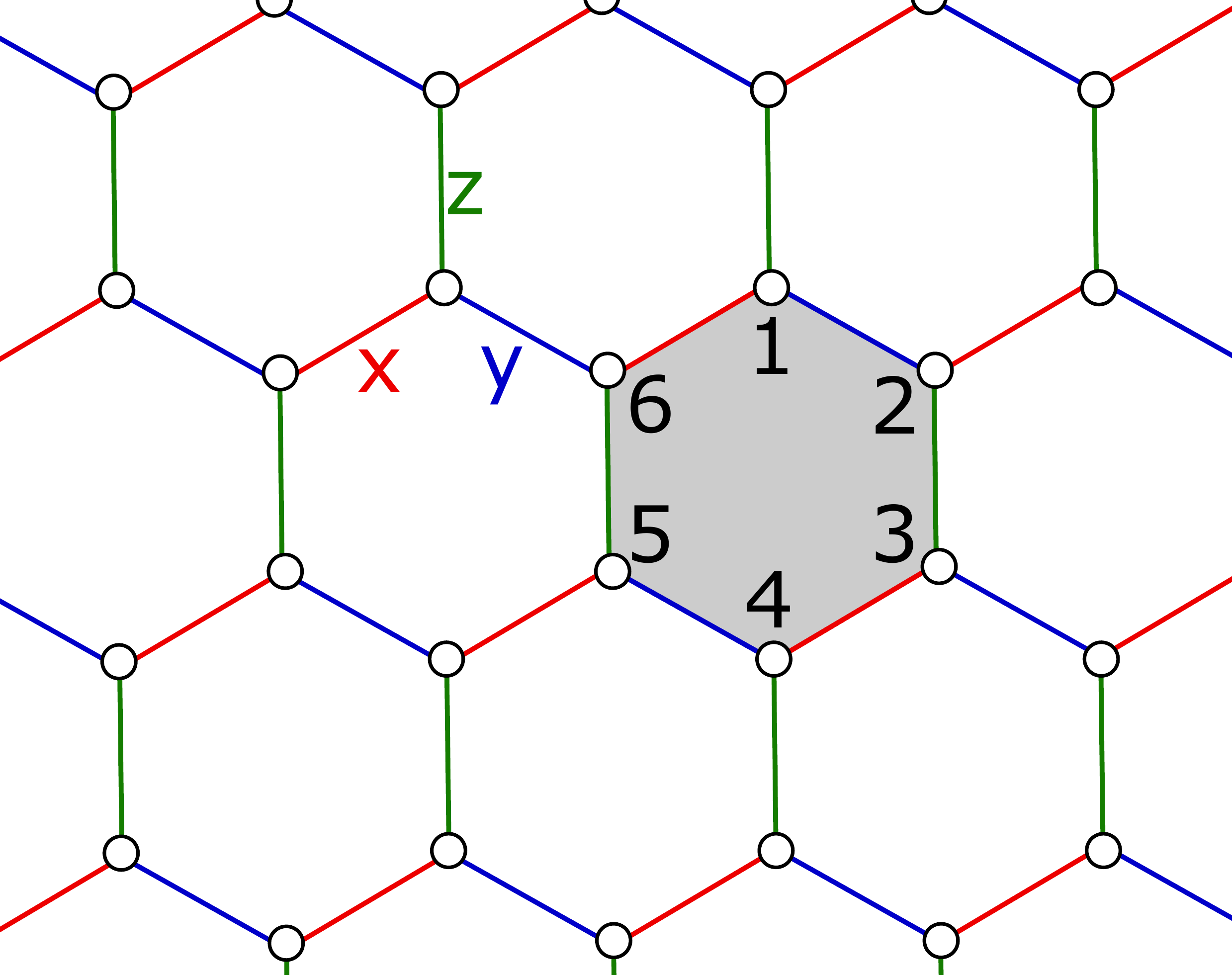}
    \caption{A honeycomb lattice showing the three kinds of interactions between neighbouring spins which are on the vertices of hexagonal plaquettes. Red, blue and green links indicate $\sigma^x \sigma^x$, $\sigma^y \sigma^y$ and $\sigma^z \sigma^z$ interactions respectively. The hexagonal plaquette operators correspond to many conserved quantities in the system. A generic plaquette whose spin sites have been numbered 1 to 6 is highlighted in grey.}
    \label{fig:honeycomb}
\end{figure}

Although diagonalization of the Kitaev honeycomb model is not required for solving the time-dependent control problem we define later, we still outline its strategy, as the operators and terminology introduced will also play a role in the time-dependent version.
The model takes its name from its hexagonal lattice geometry consisting of spin-1/2 particles located at the vertices of hexagonal plaquettes, as shown in Fig.~\ref{fig:honeycomb}.
It is defined by the Hamiltonian
\begin{align}
    H =  - \sum_{\{j,k\} \in N_2} J_s \sigma^s_j \sigma^s_k  - K \sum_{\{j,k,l\}\in N_3} \sigma^x_j \sigma^y_k \sigma^z_l,
\end{align}
where $N_2$ correspond to honeycomb edges and $N_3$ are certain triplets described further below. There are three types $s=x,y,z$ of two-body nearest neighbor Pauli interactions determined by the position of the edge in the lattice, highlighted in three colors in Fig.~\ref{fig:honeycomb}.
The three-body terms act within each hexagonal plaquette in the following way: three adjacent spins contribute to a three-body interaction term with the middle spin interacting through the Pauli operator corresponding to the link pointing outwards from the plaquette, while each of the two remaining spins interact through the Pauli operator corresponding to their link to the middle spin. For example in the plaquette highlighted in Fig.~\ref{fig:honeycomb}, one of the three-body interaction terms would be $\sigma^y_1 \sigma^x_2 \sigma^z_3$, with $5$ similar terms following clockwise along the hexagonal plaquette. While the two-body part of the Hamiltonian allows for the model to be solved by a process of Majorana fermionization, the three-body part preserves the solvability of the model while also breaking time-reversal symmetry and it consequently gives the system non-trivial topological order \cite{Kitaev2006,Pachos2012}.

For every hexagonal plaquette, a corresponding plaquette operator $W_p$ may be defined which acts on every spin with the Pauli operator of the outward pointing interaction, so that for example on the numbered gray plaquette in  Fig.~\ref{fig:honeycomb} we have $W_p = \sigma^z_1 \sigma^x_2 \sigma^y_3 \sigma^z_4 \sigma^x_5 \sigma^y_6$.
Each plaquette operator squares to the identity so that its eigenvalues adopt the values $\pm 1$ only.
Since the plaquette operators all commute with the Hamiltonian and with one another, the system Hilbert space is naturally partitioned into simultaneous eigenspaces of all plaquette operators.

Negative plaquette eigenstates are known as vortices and by a well known theorem  \cite{Lieb1994FluxBand} it is known that the ground state eigenspace is in the no-vortex sector \cite{Pachos2012}. Different vortex sectors relate to the presence of anyons localized at the respective vortex plaquettes.

While solving the system remains an intractable exponential problem even after restriction to one vortex sector, a mapping of the problem into a Majorana fermionic Hamiltonian provides a pathway towards diagonalization. For a more detailed breakdown of diagonalization see \cite{Kitaev2006,Pachos2012}. The mapped Majorana fermionic Hamiltonian is defined by replacing spin qubit sites $j$ with two fermionic sites and their corresponding creation operators $a^\dagger_{j,1}$ and $a^\dagger_{j,2}$. For each site $j$, the real and imaginary parts of the two fermionic modes constitute a total of four Majorana modes per site. The Majorana creation/annihilation operators are defined
\begin{align}
    b^x_j &= a_{j,1} + a_{j,1}^\dagger, ~~~  
    b^y_j = \frac{1}{i} (a_{j,1} - a_{j,1}^\dagger), \\
    b^z_j &= a_{j,2} + a_{j,2}^\dagger, ~~~
    c_j = \frac{1}{i} (a_{j,2} - a_{j,2}^\dagger).
\end{align}

Since with this mapping the Hilbert space is enlarged, a projection is required to obtain vectors that correspond to states in the original Hilbert space of the model. This requires the stabilizer projector \cite{Kitaev2006,Pachos2012}
\begin{align}
    P_D = \prod_j^N \left( \frac{1+D_j}{2} \right)
    \label{PDdefinition}
\end{align}
with $D_j = b^x_j b^y_j b^z_j c_j$.

The newly mapped Hamiltonian 
\begin{align}
    H_f = \frac{iJ}{2} \sum_{\{j, k\}\in N_2}  \hat{u}_{jk} c_j c_k + \frac{iK}{2} \sum_{\{j,l\}, \{k,l\} \in N_2} \hat{u}_{jl} \hat{u}_{kl}  c_j c_k  ,
    \label{fermpicture}
\end{align}
is defined in terms of Majorana operators where we may also define link operators $\hat{u}_{jk} = ib^s_j b^s_k$ in a system of $L$ total links and $N$ spin qubits.

The eigenvalues $\pm 1$ of link operators allow for further partitioning of each vortex sector into link sectors. To this end, we can define a corresponding link sector projector \begin{align}
    P_u =  \prod_{\{j,k\} \in N_2}^L \frac{1 + u_{jk} \hat{u}_{jk}}{2},
    \label{Pudefinition1} \end{align}
which amounts to a tuple of chosen eigenvalues $u_{jk}\in \{\pm 1\}$.    

Picking a certain link sector corresponds to fixing a gauge for a specific vortex sector and leads to a quadratic fermionic Hamiltonian $P_u H_f P_u$ that is easily diagonalized~\cite{Blaizot1986QuantumSystems}.  The trivial gauge would consist of setting all link eigenvalues to $u_{jk}=1$, alongside the constraints imposed by their anti-symmetry  $\hat{u}_{jk} = -\hat{u}_{kj}$. This amounts to defining an orientation which for concrete purposes we define as follows: a positive orientation on an $x$-link is directed from the bottom-left qubit to the top-right one ($j$ to $k)$, for a $y$-link it is directed from the bottom-right to the top-left, and for a $z$-link it is directed from top to bottom.

The Hamiltonian becomes diagonal in a certain quasiparticle basis $
    H = \sum_{\omega >0} \omega_j b_j^\dagger b_j - E_g
$ and the ground state is the quasiparticle vacuum state. As previously stated, for physical states of the original Hamiltonian, the states will need to be projected; thus eigenstates of the spin Hamiltonian $\ket{\Psi}$ are related to eigenstates of the quadratic fermionic Hamiltonian $\ket{\psi}$ by
\begin{align}
    \ket{\Psi} = P_D P_u \ket{\psi}.
\end{align}
The energy gap in the system corresponding to the vortex energy $\delta$ is simply the difference in ground state energy $E_g$ between differing vortex sectors. In general this is the energy gap between pairs of vortices but in exceptional examples, such as with the single plaquette lattice, this gap corresponds to a single vortex.

Additionally the fermionic excitation gap is equivalent to the smallest fermionic energy $\omega_j$ within a chosen vortex sector.

\section{Anyon creation as an optimal control problem}

Let us first describe the usual adiabatic approach. Vortex creation is implemented in terms of the time-dependent Hamiltonian
\begin{align}\label{adiab}
    H(t) &= H + \frac{t}{T} H_\text{control}^{j,k},
\end{align}
comprised of the original honeycomb Hamiltonian as a drift combined with a control Hamiltonian

\begin{align}
    H_\text{control}^{j,k} &= 2J_{jk}\sigma^s_j \sigma^s_k + 2K \sum_{j,k \in \{a,b,c\}} \sigma^x_a \sigma^y_b \sigma^z_c.
\end{align}
where $\sigma^x_a \sigma^y_b \sigma^z_c $ are those three-body terms such that two of $a,b$ and $c$ are $j$ and $k$. There are four such three-body terms for a chosen $j,k$ pair due to the definition of the three-body interactions described in the previous section.
This amounts to gradually reversing the sign of a specific $s$-link, as well as the sign of the nearby three-way interactions, using a linear time-dependence, with steepness and therefore adiabaticity determined by the duration $T$ of the adiabatic protocol. As we will see in the numerical examples, $T$ needs to be very large to obtain a good fidelity.

We now want to set up anyon creation as an optimal control problem in the hope that we can obtain similar fidelities in much shorter times compared to the adiabatic evolution. To this end, generalize the time-dependence of Eq. (\ref{adiab})  as
\begin{align}
    H(t) &= H + f(t) H_\text{control}^{j,k},
    \label{timedepH}
\end{align}
where $f(t)$ is the \emph{ramp function} defined such that $f(0)=0$ at the initial point in time $t=0$ and such that $f(T)=1$ at the final point in time $t=T$. Typically, $f(t)$ is assumed to be piecewise smooth or piecewise constant.

A typical figure of merit to be maximized is the state fidelity $\mathcal{F}$ defined in terms of an initial state $\ket{\Phi_0}$, the propagator $U[f(t),T]$ induced by the time-dependent Hamiltonian $H(t)$ and a target state that is meant to be created. In the present case, the initial state $\ket{\Phi_0}$ would usually be the ground state of the Kitaev honeycomb model whereas the target state $\ket{\Phi_\text{target}}$ is a state with an additional vortex-pair created.

While optimizing for such a state fidelity is a generally successful approach,
it has two flaws when it comes to the Kitaev honeycomb model. Since topological stability only arises for large lattice size, any practical application of the model requires a vast number of qubits. Evaluating the time-evolution operator therefore requires numerics in exponentially large spaces. Secondly, even if the ground state is solvable analytically in the free fermion picture, we would have to translate it back into the spin picture to evaluate $\mathcal{F}$, which is again exponentially hard. In the following, we will resolve both problems to obtain a scalable optimization method.

\subsection{Time-dependent fermionic picture}

Let us first describe the time-independent case. In the fermionic picture, the quadratic Majorana Hamiltonian can be written in the most general form with a matrix $J_{jk}$ that incorporates all interaction factors $J$ as $H = \frac{i}{2} \sum_{jk} J_{jk} c_j c_k$. When written in terms of full fermionic creation and annihilation operators this is:

\begin{align}
    H = \frac{1}{2} \alpha^\dagger M \alpha
    \label{hdec1}
\end{align}
where $\alpha = (a_1, \ldots, a_N, a_1^\dagger,\ldots , a_{N}^\dagger)^T$ and the Hermitian matrix $M$ 
\begin{align}
    M = \begin{pmatrix}
\mu & \nu\\
-\nu^{*} & - \mu^{*}
\end{pmatrix}
\label{Mmatrix1}
\end{align}
may be defined in terms of a Hermitian matrix $\mu$ and an antisymmetric matrix $\nu$.

A canonical transformation $T$ can then be found so that $TMT^{-1} =  \text{diag} \{\omega, -\omega\} $
where $\omega$ is a diagonal $2N$-by-$2N$ matrix. This allows for the Hamiltonian to be diagonalised in terms of quasiparticle modes   \cite{Blaizot1986QuantumSystems}.

We now consider how, in the fermionic picture, we may calculate the fidelity between a state evolved from an initialized state by a \emph{time-dependent Hamiltonian} towards a target state. To this end, we write a ground state of $H$ as $ \ket{\Phi_0}\equiv  A\ket{\text{vac}} $, with the vacuum state $\ket{\text{vac}}$ satisfying the relation $a_j\ket{\text{vac}}=0\;\forall j$. The operator $A$ is some appropriately chosen function of creation and annihilation operators. In the Appendix, we show that the state fidelity in the fermionic picture is given by
\begin{align}
    \mathcal{F}(t) &= |\bra{\Phi_\text{target}} k P_D P_u ~V[f(t),t] A \ket{\text{vac}}|^2  \\
    &= |\bra{\Phi_\text{target}} k P_D P_u A(t) V[f(t),t] \ket{\text{vac}}|^2.
\end{align}
Here the projector $P_u$ is given by Eq. (\ref{Pudefinition1}), $V[f(t),t]$ is the evolution operator corresponding to the quadratic Hamiltonian $P_u H(t) P_u$ and $P_D$ is given by Eq. (\ref{PDdefinition}), while $k$ is a real number which depends on the specific lattice (see Appendix for specific examples). 
 $A(t)$ is a Heisenberg picture operator $A(t)\equiv  V[f(t),t] A V[f(t),t]^\dagger$.

In analogy to Eqs (\ref{hdec1}) and (\ref{Mmatrix1}) it is useful to decompose $P_uH(t)P_u$ as
\begin{align}
    P_u H(t) P_u = P_u \frac{1}{2} \alpha^\dagger M(t) \alpha
    \label{alphaMalpha}
\end{align}

Since $A$ depends on annihilation and creation operators, we may write it as $A(\alpha)$. It can then be shown \cite{Blaizot1986QuantumSystems} that

\begin{align}
    V[f(t),t] A(\alpha) V^{\dagger} (t) = A( W[f(t),t] \alpha) 
\end{align}
where $W(t)$ is the $2N$-by-$2N$  the time-ordered product solving the differential equation
 \begin{equation}\label{heisenberg}
 	\dot{W}[f(t),t]=iM[f(t)]W[f(t),t].
 \end{equation}

This solves the problem of an exponentially sized evolution operator, as $W(t)$ scales linearly in the system size. We will refer to calculations in this  picture as the  Heisenberg picture, since it is directly obtained from the Heisenberg equations of motions of $\alpha$.  However, the problem of expressing the target and initial state in the spin picture remains. This will be tackled next.

\subsection{Heisenberg fidelity as optimization target}
In the previous paragraph we showed that the evolution is fully determined by the Heisenberg picture of the quadratic Hamiltonian. If we knew a good target evolution, rather than target state, we could therefore free ourselves from the state picture and obtain all quantities directly in the Heisenberg picture.
The key idea here is to get back to the adiabatic evolution to find such good evolution. We phrase such evolution directly in the Heisenberg picture. To this end, let $W_{\text{ad}}$ be the solution of Eq. (\ref{heisenberg}) for the adiabatic ramp Eq. (\ref{adiab}) with some suitably large duration $T_\text{ad}$. This can be computed efficiently without having to refer to states. We define a corresponding Heisenberg fidelity
\begin{align}
    \mathcal{F}_H = \frac{1}{2N} \left| \Tr (W_\text{ad}^\dagger W[f(t),T]) \right|
\end{align}
This quantity obtains its maximum of $1$ if and only if the evolutions match up to a phase and it can be used for efficient numerical optimization. In Appendix B, we show that 
\begin{align}
    1-\mathcal{F}_H \geq \frac{1}{32N^3}\left(1-\sqrt{\mathcal{F}}\right).
    \label{bound}
\end{align}
This shows that $\mathcal{F}_H$ is a good surrogate for $\mathcal{F}$ and may be optimized instead. In order to appropriately compare very high fidelities and plot them logarithmically we give our results in terms of infidelity $\mathcal{I}_H = 1- \mathcal{F}_H$.

\section{Numerical results}
Here, we use the QuTiP \cite{Johansson2013QuTiPSystems} implementation of the gradient-ascent pulse engineering (GRAPE) algorithm \cite{Khaneja2005OptimalAlgorithms,Rowland2012ImplementingPracticalities} using the limited memory version of the Broyden–Fletcher–Goldfarb–Shanno method (BFGS) as an optimizer with exact gradients \cite{Broyden1970TheConsiderations,Fletcher1970AAlgorithms,Goldfarb1970AMeans,Shanno1970ConditioningMinimization}. The optimization takes place over piecewise-constant functions, which means that the number of time-steps becomes an additional parameter of our numerics. This algorithm begins with a random choice of initial pulse. Using BFGS, exact gradients of the infidelity are calculated and parameters within the pulse are varied as the infidelity goes 
down in the optimization space. This is done by calculating second derivatives of the infidelity with respect to the parameters, using this to approximate a parabola, and descending to the minima of this approximation until reaching a tolerable minimum infidelity. The limited memory version of this algorithm (L-BFGS) does not require calculation of the entire Hessian matrix of second derivatives and so is not computationally intensive \cite{Liu1989OnOptimization}.

\subsection{Optimized non-adiabatic pulse in a simple lattice}

While the timescale of anyon creation through adiabatic evolution can be very long, if instead of using linear ramps we use non-linear time-dependence in the Hamiltonian which have been specially designed, then we can achieve high fidelities at shorter timescales. The well-tested aforementioned GRAPE algorithm is used to develop such time-dependent control functions also known as pulses. The explicit Hamiltonians used in our optimizations incorporate interaction factors $J$ of magnitude 1 with the effective magnetic field interaction term equaling $J/100$, and thus all times are considered to be in units of $J^{-1}$ to be applicable to the general case.

\begin{figure}[h]
    \centering
        \includegraphics[width=\linewidth]{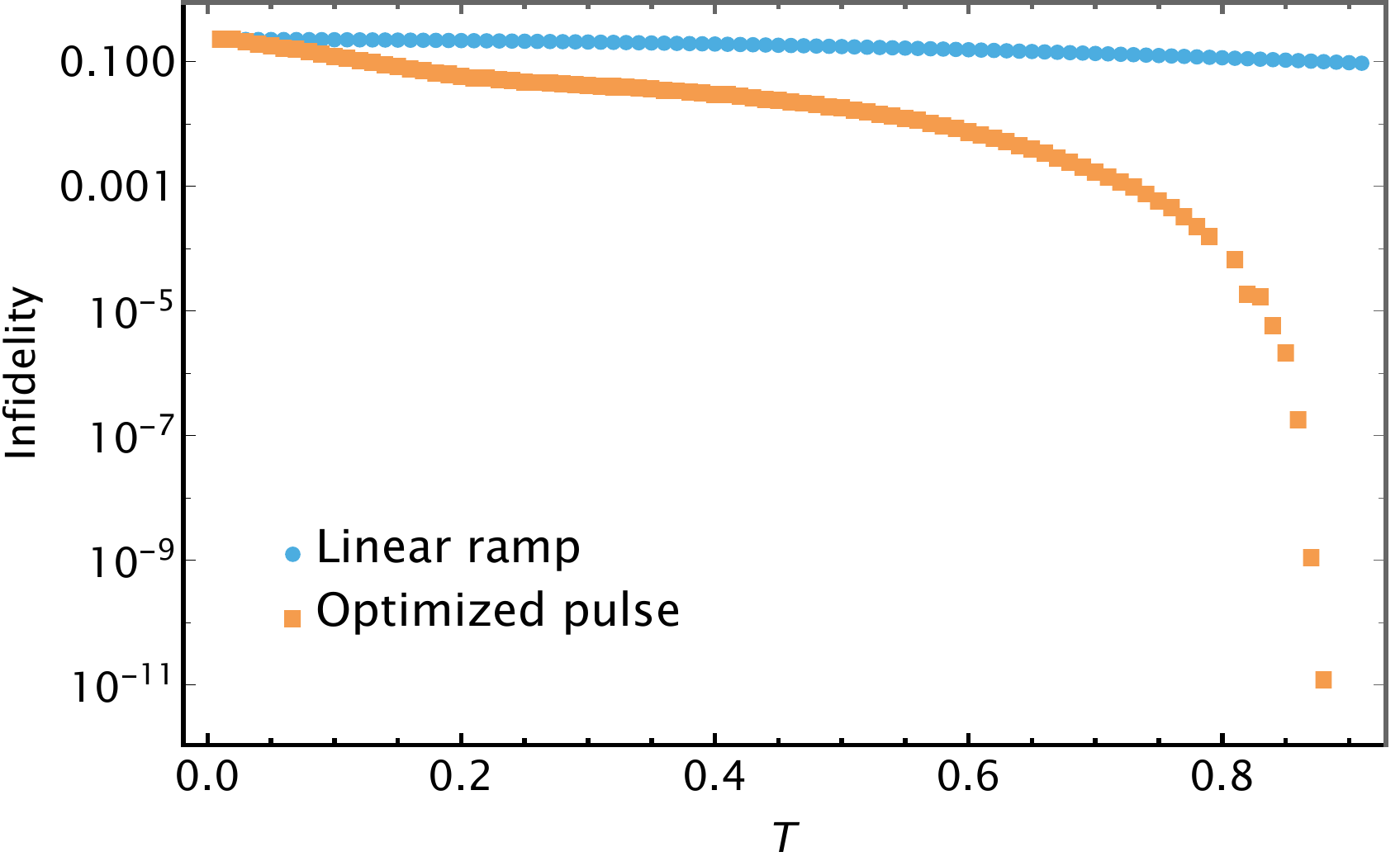}
    \caption{Logarithmic comparison of infidelity between initial and adiabatic target state for linear control pulses and optimized control pulses, at various timescales, for the simplest 6-qubit lattice. 100 time steps were used in both cases. We can see more clearly the dramatic improvement in fidelity at approximately $T=0.8$. The minimum infidelity reached by optimized pulses, on the order of $10^{-9}$ is reached at time $T \approx 0.9$, many orders of magnitude less than the time to reach this infidelity with the linear ramp, which is at  $T \approx 1350$.}
    \label{fig:qutipresults}
\end{figure}

The results of using this procedure for a single plaquette of 6 spins are depicted in Fig.~\ref{fig:qutipresults} showing infidelities as function of the ramp time $T$.
The infidelities obtained with a linear ramp are depicted in blue.
There is a slight improvement with increasing ramp time, but the fidelity of about $90\%$ achieved with a ramp time of $T=1$ is only a very small improvement compared to the initial fidelity at time $T=0$.
This is consistent with an estimate based on the spectral gap condition \cite{VanDam2001HowComputation} that implies ramp times $T \gg {\delta}^{-2}= 3.480\ldots$ are required for high-fidelity operations. With a value of $J=1$ this spectral gap is calculated from the vortex gap of $\delta =| -4J-(-2\sqrt{3}J)| \approx 0.536$. This is a narrower gap than the first fermionic excitation energy penalty which is $2J$. While in general the fermionic excitation gap is a function of $J$, and $K$ and so can be narrower than the vortex gap, in this special case of the single plaquette lattice this is not the case. 

The behaviour with optimized ramps, depicted in orange, is fundamentally different.
In the range $T<0.2$ there is a much faster decrease of infidelity with increasing ramp time than in the case of linear ramps.
This decrease is a bit less pronounced in the range $0.2\lesssim T\lesssim 0.4$,
but for ramp time $T>0.4$ this decrease becomes increasingly pronounced with increasing ramp time.
For ramp times $T\gtrsim  0.8$, there is a rapid drop in the infidelity, and for ramp times exceeding the threshold value of $T=T_d\simeq 0.85$,
the deviations between the numerically obtained infidelities and the ideal value of $0$ are consistent with noise due to finite numerical accuracy.

It is by no means surprising that even with an optimized ramp a finite ramp time is required to reach perfect fidelities.
This is due to the constant part in the system Hamiltonian Eq.~(\ref{timedepH}) that defines a natural time-scale of the system.
This effect is also referred to as the {\it quantum speed limit} \cite{Burgarth2022QuantumLimits,Deffner2017QuantumControl} and we will refer to the threshold value $T_d$ of ramp durations at which fidelities drop to values close to their ideal value as the {\em drop time}.
Apart from limitations imposed by finite-dimensional parametrization of the ramp function, the numerical optimization routine and  numerical accuracy, this drop time coincides with the minimal duration required to reach perfect fidelity.
 
\begin{figure}
        \centering
        \includegraphics[width=\linewidth]{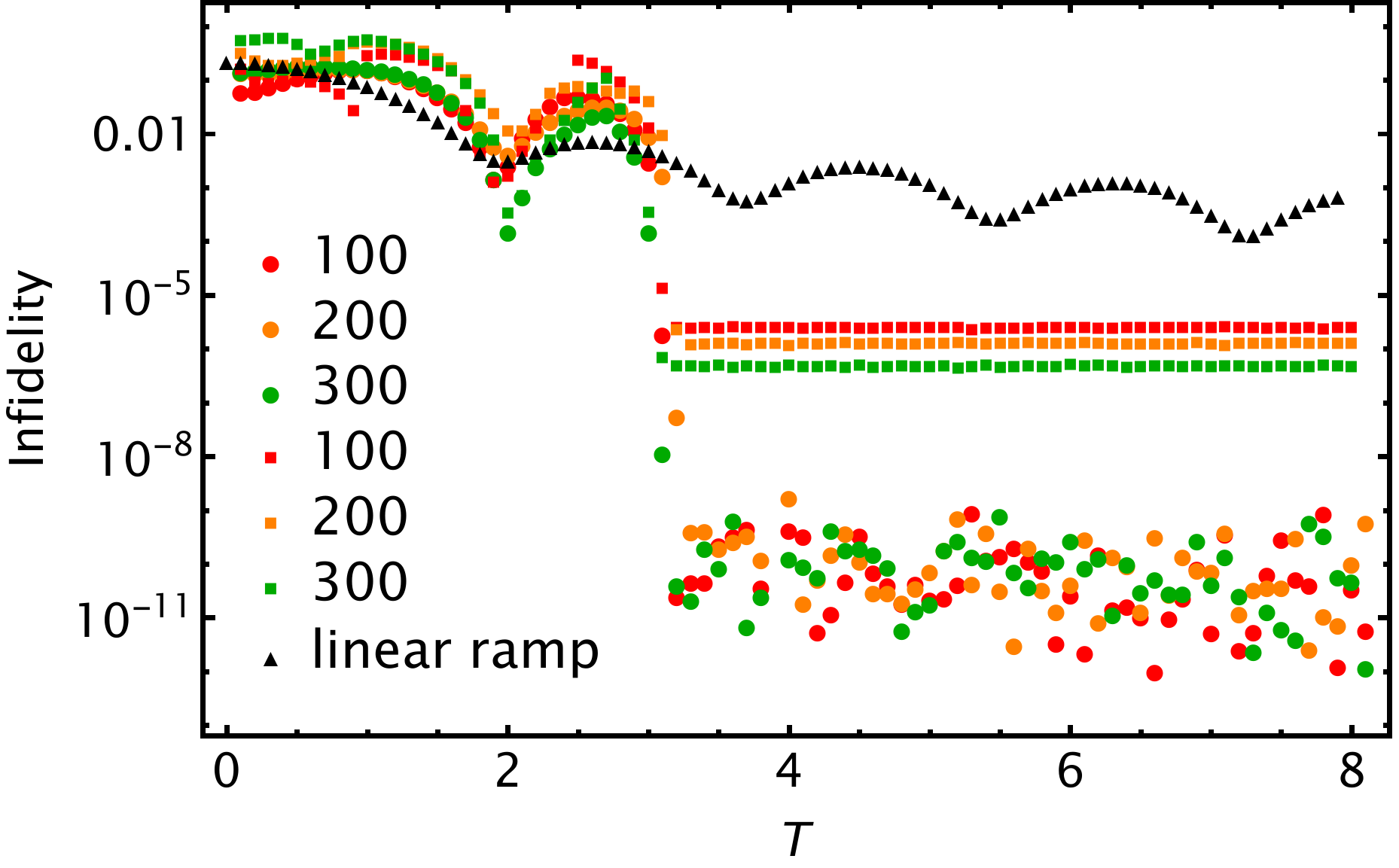}
        \caption{\textbf{6-qubit lattice}: Heisenberg infidelities based on differing targets (circles) are compared with their corresponding state fidelities (squares). These are compared overall with state fidelities achieved from linear ramp pulses (triangles). Different colors represent different values of $T_\text{ad}$, which are the timescales of 100 (red, e.g. upper squares), 200 (orange, middle squares) and 300 (green, lower squares) that define the target unitary $W_\text{ad}$. All optimized pulses are comprised of $200$ time steps. There is a significant improvement at times between $T=2$ and $T=4$. Thereafter there is effectively perfect fidelity with the presence of numerical noise. It confirms that both Heisenberg and state infidelities improve markedly at the same time $T$ and this is an improvement on the fidelities achieved with a linear ramp pulse.}
        \label{fig:6qbtresults}
\end{figure}

The example of a single plaquette with $6$ qubits is also a good test-case to compare optimization of state-fidelity and Heisenberg fidelity.
Fig.~\ref{fig:6qbtresults} depicts the Heisenberg fidelity obtained with various ramp functions as function of the ramp time $T$.

The black triangles represent state fidelity data obtained with linear ramps. Consistently with Fig.~\ref{fig:qutipresults}, there is only a moderate decrease of the infidelity with increasing ramp time.
The circles correspond to ramp functions optimized for Heisenberg fidelity, and the different colors correspond to different chosen adiabatic target times with $T = 100$, $200$ and $300$ for red, orange and green respectively.
Similarly to the observations in Fig.~\ref{fig:qutipresults} there is a clear drop of the infidelities at a drop time $T_d\simeq 3$.
The fact that the numerically observed drop time is essentially the same in all three cases indicates that the drop time is not dependent on the length of the adiabatic target time that is chosen.

The squares depict the state infidelity obtained from the ramp functions that had been optimized for Heisenberg infidelity, in order to see numerical evidence that when implementing our procedure, good fidelity is achieved in the one case which ensures a good infidelity in the other.

Here also is a clear drop of the infidelities and it occurs at the same drop time as for Heisenberg infidelities.
The fidelities obtained for ramps with a longer ramp time than the drop time $T_d$, however, are not merely limited by numerical accuracy, but they are indeed finite.
Their exact value depends on the parametrization of the ramp function, with finer parametrizations resulting in lower infidelities.
Since state fidelity and Heisenberg fidelity are not strictly equivalent, it is not surprising, that a ramp that is optimized for one of these fidelities does not yield the optimally achievable value of the other fidelity.
The results in Fig.~\ref{fig:qutipresults}, however, clearly show that ramps optimized for Heisenberg fidelity result in high state fidelities and, in particular, in infidelities that are between $3$ and $4$ orders of magnitude lower than infidelities obtained with linear ramps.

\begin{figure}
        \centering
        \includegraphics[width=\linewidth]{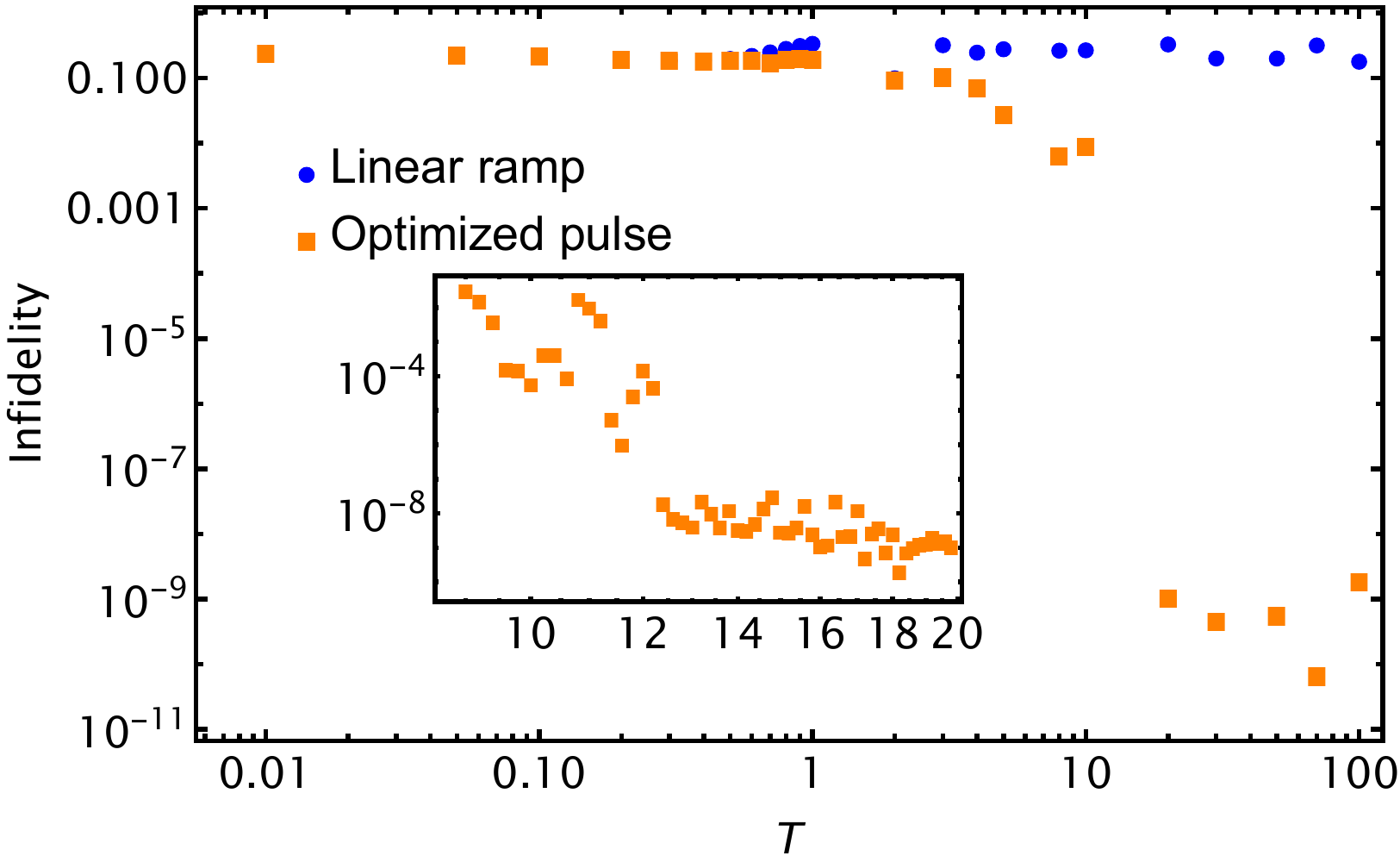}
        \caption{\textbf{10-qubit lattice}: Heisenberg infidelities between optimized unitaries and an optimized target are shown (orange), as compared with infidelities between the target and a unitary defined by a linear ramp control (blue). Each optimized pulse consists of 200 time steps. The drop time is $T \approx 10$, later than for the 6-qubit lattice.}
        \label{fig:10qbtresults}
\end{figure}

\begin{figure}
        \centering
        \includegraphics[width=\linewidth]{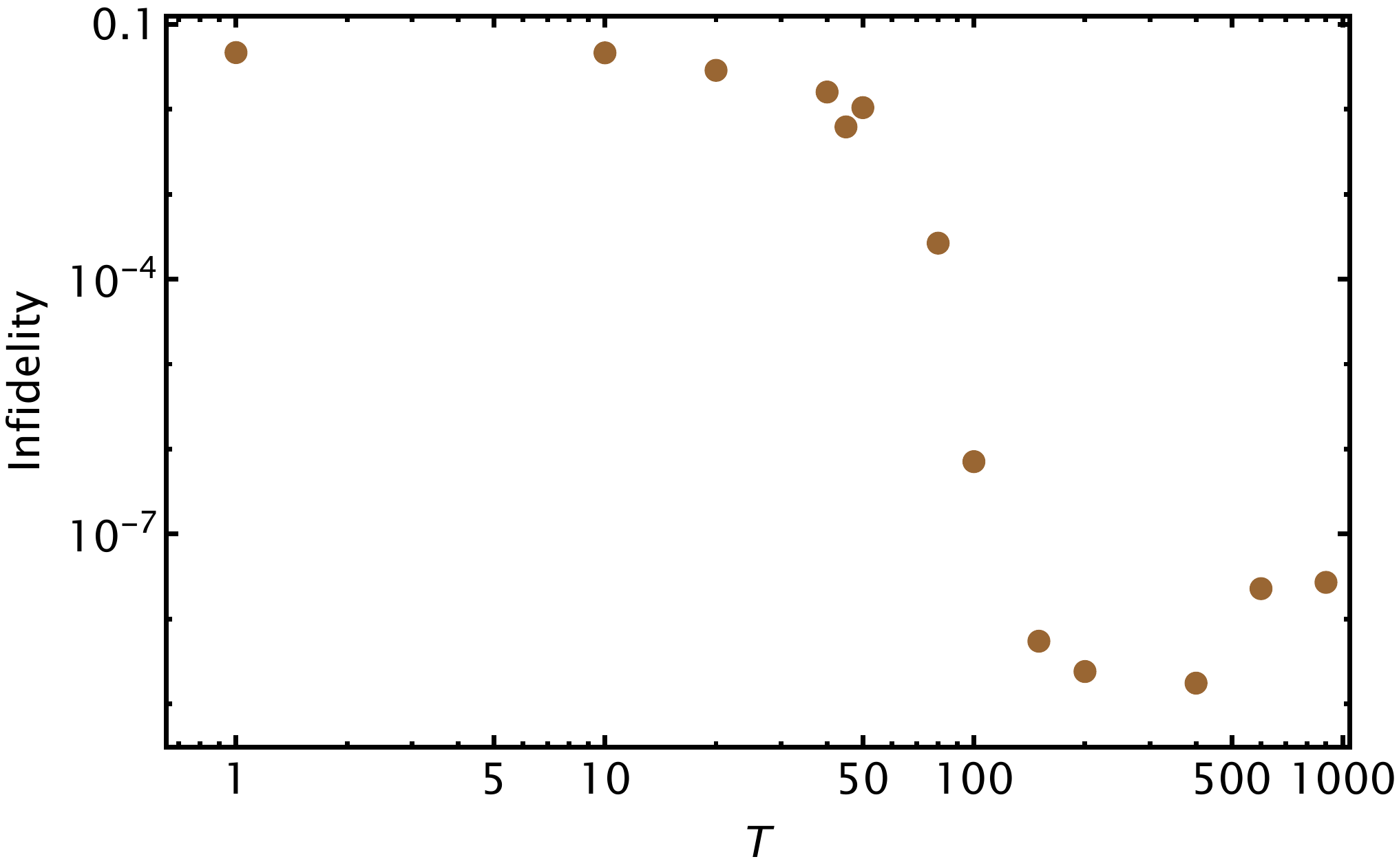}
        \caption{30-qubit lattice with open boundary conditions and pulses with 400 time steps. Heisenberg infidelities between optimized unitaries and an optimized target are shown. The spectral gap condition requires that $T_\text{ad} \gg 30$ calculated analogously to the 6 and 10-qubit lattices. The results show that $T_d$ is within the order of magnitude of 30 rather than many orders of magnitude larger as required by adiabaticity, highlighting the improvement in timescale achieved.}
        \label{fig:30qbtresults}
\end{figure}

\subsection{Optimized pulses for larger lattices}

Since numerical optimizations of state fidelity rapidly become infeasible with increasing system size, the subsequent examples for larger systems feature only Heisenberg fidelities with ramps that are optimized for this Heisenberg fidelity.
Fig.~\ref{fig:10qbtresults} shows the Heisenberg infidelity as function of ramp time for linear ramps (blue) and for optimized ramps (orange) for a lattice consisting of two adjacent plaquettes made up of 10-spins.
Similarly to the cases discussed above, there is a clearly identifiable drop time $T_d$, but its value $T_d\simeq 10$ is larger than in the examples of smaller systems. 
The abscissa depicts that ramp time on the log-scale, highlighting that linear ramps with durations exceeding the drop time by many orders of magnitude are required to achieve any sizable decrease in infidelity. The specific adiabatic timescale is a consequence of the energy gap between vortex pairs. This is calculated in the same way as for the single-plaquette system and with the parameters being set at $J=1$ and $K=J/100$ this gap is $\delta = 0.375\ldots$, requiring a timescale of $T \gg \delta^{-2} \approx 7.070$.

The inset depicts a zoomed-in look into the domain around the drop time.
It highlights that, on top of the rapid drop of infidelity there is also a finite noise level. When we compare the optimization results of a system made up of 10 spin-qubits and one with 30-qubits, whose optimized infidelities are shown in Fig.~\ref{fig:30qbtresults}, we see again the marked increase in drop time that is achieved. The geometry of the 30-qubit spin lattice is identical to the lattice systems presented in Fig.~\ref{fig:ylinksandzlinksflipped} and is subject to open boundary conditions. As detailed in Appendix A, lattices with periodic boundary conditions require up to four times as much optimization to be carried out compared with the same geometry without periodicity therefore showing results for open boundary conditions is the most efficient choice given limited computational resources. Here the spectral gap condition, calculated analogously as with the 6 and 10-qubit systems, requires the adiabatic timescale for this system to be $T \gg \delta^{-2} \approx 30$. In the $30$-qubit lattice the vortex gap remains smaller than the lowest fermionic energy gap but for larger systems this may not necessarily 
be the case. Additionally the smallest gap may not be the vortex gap as compared with the fermionic gap in scenarios where vortices are created from a non-zero vortex sector. 

\begin{figure}[h]
\begin{subfigure}{.5\textwidth}
  \centering
  % include first image
  \includegraphics[width=\linewidth]{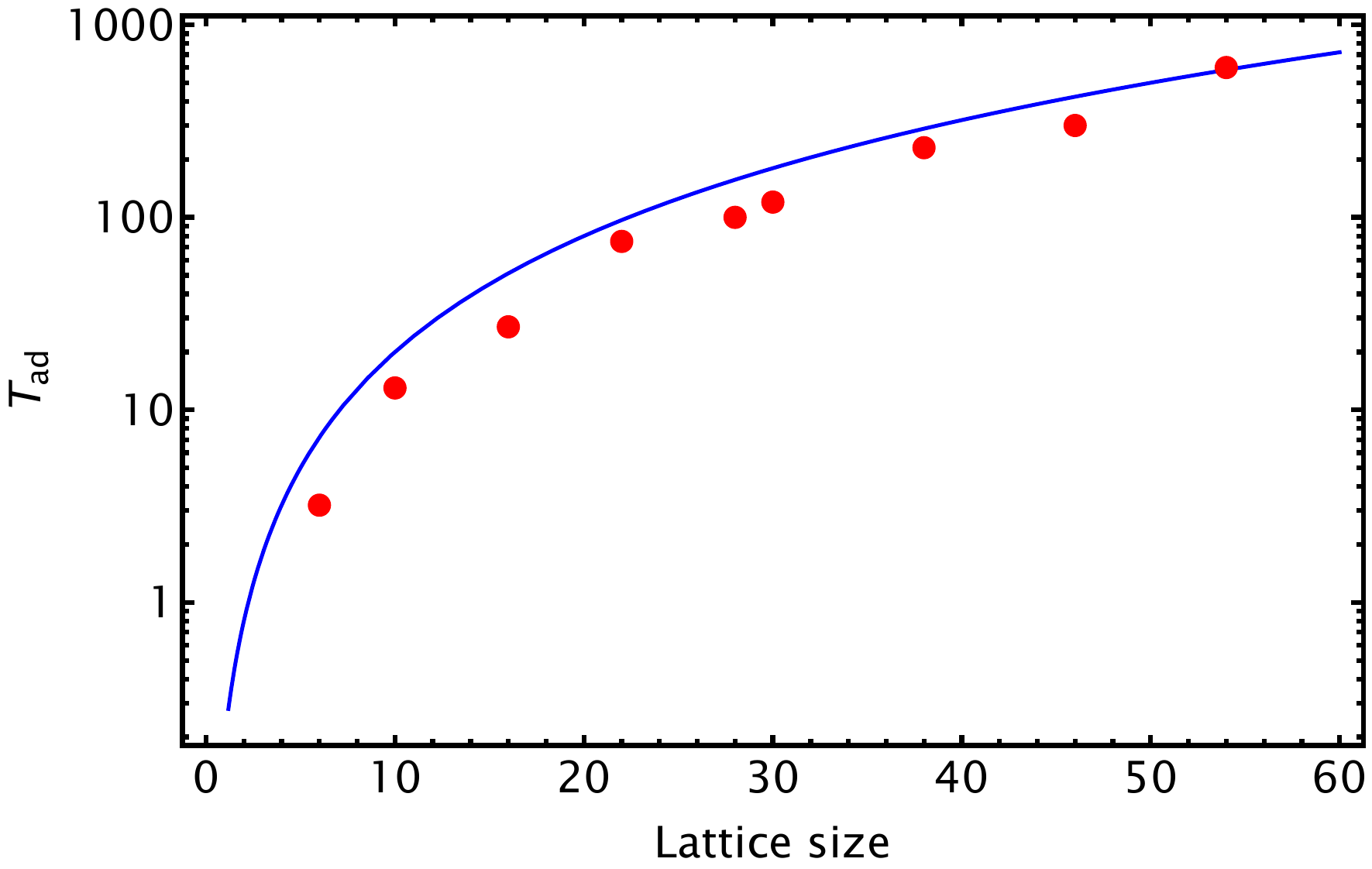}
  \label{fig:egot6}
\end{subfigure}
\begin{subfigure}{.5\textwidth}
  \centering
  % include second image
  \includegraphics[width=\linewidth]{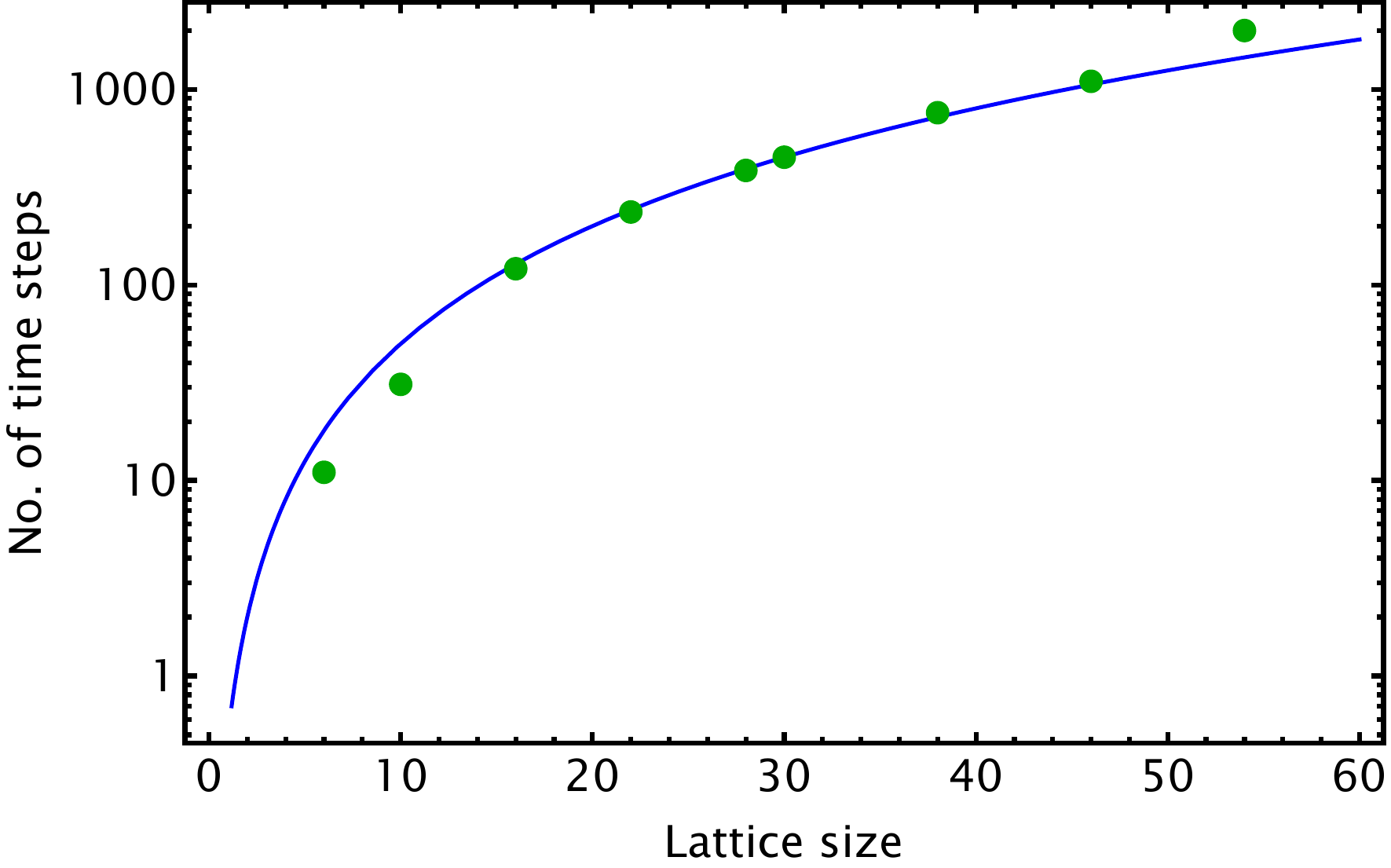}
  \label{fig:minimumNTSgraph6}
\end{subfigure}
	\caption{Drop time (\textbf{Above}) and minimum required time steps (\textbf{Below}) compared with lattice size by number of qubits. These quantities are defined as the earliest time $T$ and minimum number of time steps in an optimized piecewise-constant pulse respectively where Heisenberg infidelity drops below $10^{-6}$. They are compared with quadratic growth curves (blue) to show sub-exponential growth.}
	\label{fig:EGOT}
\end{figure}

The observation that the drop time increases with the system size is depicted more systematically in Fig.~\ref{fig:EGOT}.
Generally, the practically achievable drop time depends on the parametrization of the ramp function and the number of free parameters that can be optimized.
Only for a ramp function with sufficiently many piecewise constant elements is the drop time independent of the number of free parameters.
Fig.~\ref{fig:EGOT} depicts that shortest achievable drop time together with the minimal number of piecewise constant elements in the ramp function that is required to achieve this drop time.
In both sub-plots the scaling is consistent with a quadratic increase (blue line) with system size.

Within the validity of extrapolation from numerically accessible system sizes, there is thus a very moderate scaling with system size, highlighting that numerical optimizations based on the framework of fermionization can be performed efficiently.

\section{Conclusions}

It is well known that the Kitaev honeycomb model is a useful system for the encoding of quantum information using non-abelian anyons. Our use of quantum control techniques has allowed us to create anyons not only in adiabatic timescales, but to overcome the necessity for such a restriction through the use of gradient ascent pulse-engineering to construct non-linear ramps whose timescales are many orders of magnitude faster than adiabatic. These ramps achieve very high fidelities at these much faster timescales, and thus are more useful for the encoding of quantum information and quantum algorithm implementation given that decoherence times entail a preference for shorter timescales.

Along with the drawback of requiring long timescales with adiabatic control, the other chief drawback we would encounter with non-adiabatic quantum control is the difficulty in completely determining the dynamics of honeycomb lattices with large numbers of spin qubits. The method we have developed allows us to overcome this obstacle
by solving a matrix control problem where the matrices grow linearly in the number of lattice spin-qubits, allowing for the implementation of non-linear pulse-engineering and optimization of a related matrix or Heisenberg fidelity. This, alongside the analytic expression directly comparing state fidelity and Heisenberg fidelity allows us to be assured of the success of these optimized ramps and allows us to have confidence in the successful implementation of our procedure for use in experimental realization of the model. Given the various avenues suggested for experimental implementation of the Kitaev honeycomb model such as in solid state materials \cite{Takagi2019ConceptLiquids} and with polar molecules in optical lattices \cite{Micheli2006AMolecules}, as well as the success in observing topological order in quantum spin liquids using the Rydberg blockade mechanism \cite{Verresen2021PredictionBlockade,Semeghini2021ProbingSimulator} we believe our control methods would be a useful for realization within a setup that allows for varying spin-qubit interactions in a time-dependent manner. 

Furthermore we have seen that there is indeed sub-exponential growth in the computational difficulty of our control problem with growing system size and so carrying out our methods on lattices on the order of hundreds of qubits is possible, as necessary for scalable topological robustness.

\section*{Acknowledgements}
OR acknowledges funding from EPSRC Quantum Systems Engineering Hub and support from MQCQE. DB acknowledges funding by the Australian Research Council (project numbers FT190100106, DP210101367, CE170100009). Financial support from the QuantERA ERANET Cofund in Quantum Technologies implemented within the European Union’s Horizon 2020 Programme under the project Theory-Blind Quantum Control TheBlinQC and from EPSRC under the grant EP/R044082/1 is gratefully acknowledged

\begin{appendix}

\section{Dynamics of the time-dependent Hamiltonian in the fermionic picture}

Previous work on the Kitaev honeycomb model has primarily been focused on time-independent Hamiltonians, yet for quantum control we require time-dependent Hamiltonians. Examples of such investigations of a time-dependent Kitaev honeycomb model have included those focused on periodic driving  \cite{Fulga2019TopologyModel,Molignini2021CrossdimensionalModels} which have used Floquet theory, allowing for periodic driving to be mapped to an effectively time-independent system \cite{Verdeny2016Quasi-periodicallySystems}. Other studies have used Jordan-Wigner transformations without projections to look at specific time-dependent behavior such as the Kitaev honeycomb model with a quenched magnetic field \cite{Nasu2019NonequilibriumLiquids}. 

Here we show that in regular lattices with open, periodic or half-periodic lattices, the fermionization procedure with projections is still possible in the time-dependent case.

First let us recall the projections onto a link sector $u$
\begin{align}
    P_u =  \prod_{\{jk\}}^L \frac{1 + u_{jk} \hat{u}_{jk}}{2},
    \label{Pudefinition}
\end{align}
defined by an $L$-tuple of link eigenvalues $ u_{jk}\in \{ \pm 1\}$; 
the projector onto a vortex sector $w$
\begin{align}
    P_w = \prod_j^P \frac{1 + w_j \hat{w}_j}{2}
\end{align}
defined by a $P$-tuple of plaquette eigenvalues $w_j \in\{ \pm 1 \}$; and the projector onto the physical subspace of the fermionic space given by
\begin{align}
    P_D = \prod_j^N \frac{1+D_j}{2}\ ,\mbox{ with }\ D_j = b^x_j b^y_j b^z_j c_j.
\end{align}
Since
\begin{align}
    \hat{w}_j = \prod_{\{k,l\} \in w_\beta } \hat{u}_{kl},
    \label{wplaquettefermionic}
\end{align}
$P_u$ and $P_w$ commute, and moreover $w$ is fully determined by $u$. We denote this relationship as $w=\omega (u)$ and thus have $P_w P_{u} = \delta_{w, \omega(u)} P_u$ and
\begin{align}
        P_w = 
      P_w \sum_u P_u = \sum_{u : \omega(u) = w} P_{u}.
    \end{align}
$P_w$ commutes with $P_D$, $P_u$ and the time-dependent fermionic Hamiltonian $H_f(t)$, so it will suffice to restrict ourselves to a single plaquette sector $w$. While $P_D$ is the projection that determines physicality, $P_u$ will turn the fermionized Hamiltonian into a quadratic (and thereby easy to solve) one. A difficulty arises from the fact that $P_D$ \emph{does not} commute with $P_u$. We can however find another useful relationship between these projectors. 
Let $N$ be the number of qubits of the original spin lattice and $\{\Gamma_k| k=1,\ldots,2^N\}$ be the set of all possible products of stabilizer operators $D_i$ on the qubits, without repetition, where we take an arbitrary but fixed order. For our fixed $w$ consider the corresponding pre-image $\omega^{-1}(w)$ of link sectors. We define an equivalence relationship $\sim$ on this set by $u\sim v:\Leftrightarrow \exists k:P_u=\Gamma_k P_v \Gamma_k.$ Let $\kappa$ be the number of equivalence classes. Let us choose an arbitrary but fixed set of representatives $u_1,\ldots,u_\kappa$ and define $P_{\bar{u}} =\sum_{i=1}^\kappa P_{u_i}.$

Upon expanding $P_D$ in terms of stabilizers, we obtain
\begin{align}P_DP_{\bar{u}} P_D=\frac{1}{2^N}P_D \sum_{i=1}^\kappa \sum_{k=1}^{2^N}\Gamma_k P_{u_i}\Gamma_k.\end{align} To understand the right hand side better, we make a counting argument. Firstly, it follows from the anticommutation relationships between link operators and the $D_j$ that the $\Gamma_k P_{u_i}\Gamma_k$ are again link projectors. Since we sum over all $\Gamma_k$ and by definition of the equivalence classes, we know that every $u\in \omega^{-1}(w)$ appears at least once on the right hand side, and that there are no overlaps between the classes of fixed $i$. Furthermore the equality $\Gamma P_u \Gamma = P_u$ holds if and only if $\Gamma=1$  or $\Gamma=\prod_i D_i$. Therefore, there are $2^{N-1}$ distinct projectors for each $i$. It follows that 

\begin{align}P_DP_{\bar{u}} P_D=\frac{1}{2^{N}}2P_D\sum_{u:\omega(u)=w} P_u=\frac{1}{2^{N-1}}P_DP_w.\end{align}
From the commutativity relation $[H_f(t),P_D]=0$ it also follows that 
\begin{align}V(t) P_DP_w=2^{N-1}VP_DP_{\bar{u}}P_D=2^{N-1} P_DP_{\bar{u}}V(t) P_{\bar{u}}P_D,\end{align} where $V(t)$ is the propagator corresponding to $H_f(t)$. Hence, the evolution can be computed in the subspace $P_{\bar{u}}$. To conclude the argument, we need to know the value of $\kappa$, as this determines how many link sectors we need to consider. As long as $\kappa$ is not exponential, we can efficiently simulate the dynamics.

 To this end, we need another counting argument. To simplify the analysis, we only consider three different regular lattice types dubbed open (o), periodic (p) and half-periodic (h), and find their corresponding values of $\kappa$.

To do this, we first find relationships for the number of link operators $L$, the number of plaquettes $P$, and the number of qubits $N$ for the various lattices. Simple but rather tedious counting of such regular lattices shows  that $L-P=N-1$ in the open case and $L-P=N$ in the other two cases. Next, compute the size of $\omega^{-1}(w).$ We show in the lemma below that $|\omega^{-1}(w)|_{o,h}=2^{L-P}$ and $|\omega^{-1}(w)|_{p}=2^{L-P+1}.$ Since each equivalence class has exactly $2^{N-1}$ elements, we have to have 
\[|\omega^{-1}(w)|=2^{N-1}\kappa \] such that $\kappa_o=1$, $\kappa_h=2$,$\kappa_p=4$.

\begin{lemma}
\label{lemma2Lminusp}

For all $w$, $|\omega^{-1}(w)|_{o,h}=2^{L-P}$ and $|\omega^{-1}(w)|_{p}=2^{L-P+1}.$ 
\begin{proof}

Consider first the case of a lattice with open boundary conditions. Since there are no boundary constraints, all possible configuration of plaquette eigenvalues $\{\Lambda_j \}$ are possible, and
\begin{align}
    |\{ \Lambda _j \}|_{o} = 2^P.
    \label{numberofWis2p}
\end{align}
 From the above, for each $u$, we can find $2^{N-1}$ other $v$ with $\omega (v)=\omega (u)$ by conjugation with $\Gamma_k$. Since in the open lattice the number of qubits $N$ follows the relation $N-1=L-P$, once can conclude that for each $w$, the inequality $|\omega^{-1}(w)|_{o}\ge 2^{L-P}$ holds. Since there are by definition $2^L$ different link sectors, we have 

\begin{align}
2^L=\sum_w  |\omega^{-1}(w)|_{o}\ge 2^{L-P}2^{P}=2^L
\end{align} so equality holds and the statement follows.

Next, consider a lattice with full periodic boundary conditions. Any link eigenvalue change leads necessarily to exactly two plaquette eigenvalues being flipped. Therefore only even numbers of vortices may ever be present, and the number of plaquette eigenvalue configurations is
\begin{align}
    |\{ \Lambda_j \}|_p = 2^{P-1}.
\end{align}
Now, for each $u$, we can find $2^{N-1}$ other $v$ with $\omega (v)=\omega (u)$ by conjugation with $\Gamma_k$, but for each of these we can find 4 inequivalent link sectors. Since $L-P=N$ for periodic boundaries,  we obtain
\begin{align}
2^L=\sum_w  |\omega^{-1}(w)|_{p}\ge  4\times  2^{L-P-1}2^{P-1}=2^L
\end{align}
and the statement follows again.

Finally, in the half-periodic case we only have 2 inequivalent link sectors, but $|\{ \Lambda_j \}|_p = 2^{P}$ is twice as big as in the periodic case, and the same counting argument holds.

\end{proof}
\end{lemma}

\subsubsection*{Inequivalent sectors: Periodic boundary conditions}

\begin{figure}[h]
\begin{subfigure}{.5\textwidth}
  \centering
  % include first image
  \includegraphics[width=.5\linewidth]{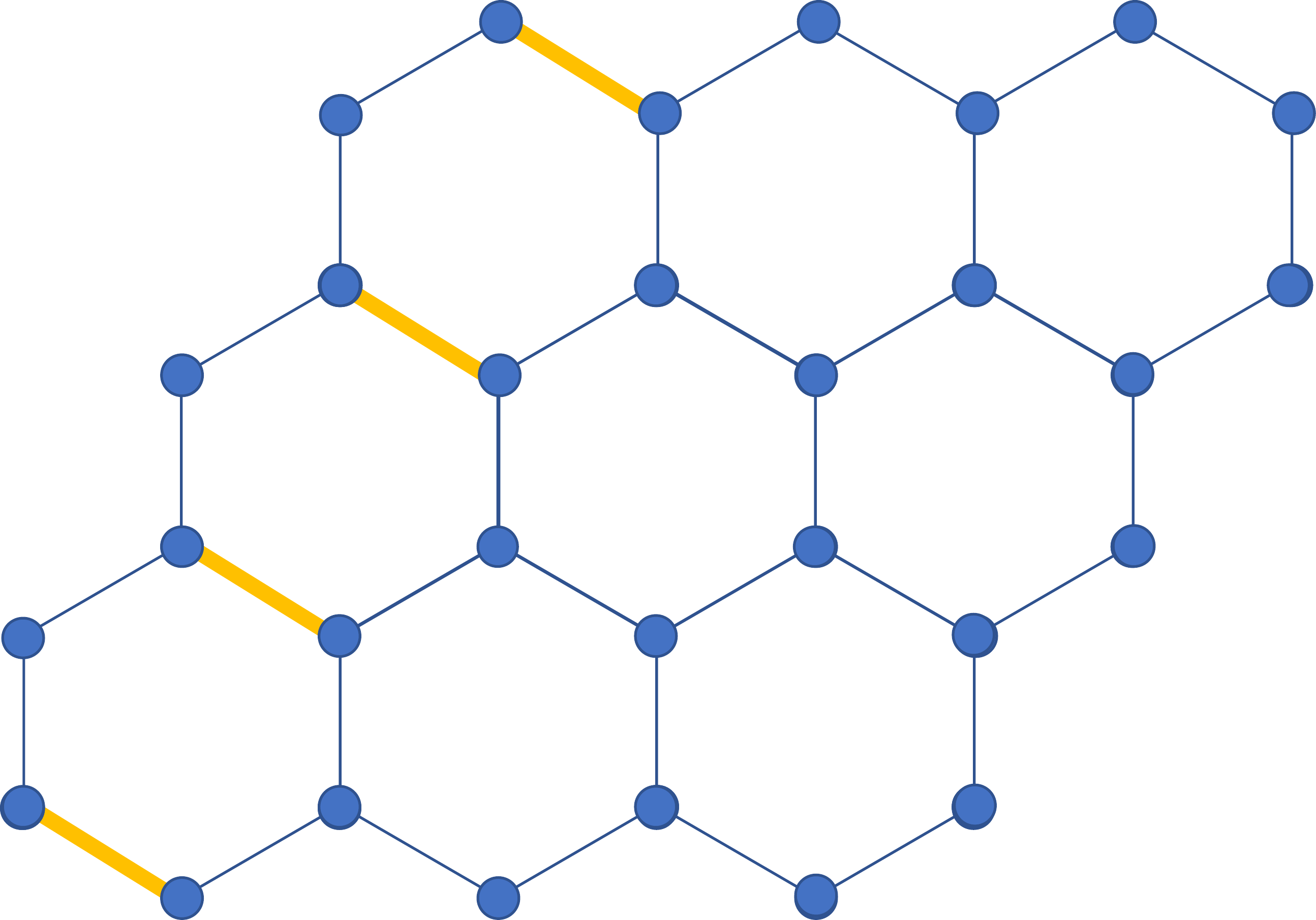}
\caption{}
  \label{fig:ylinksflipped}
\end{subfigure}
\begin{subfigure}{.5\textwidth}
  \centering
  % include second image
  \includegraphics[width=.5\linewidth]{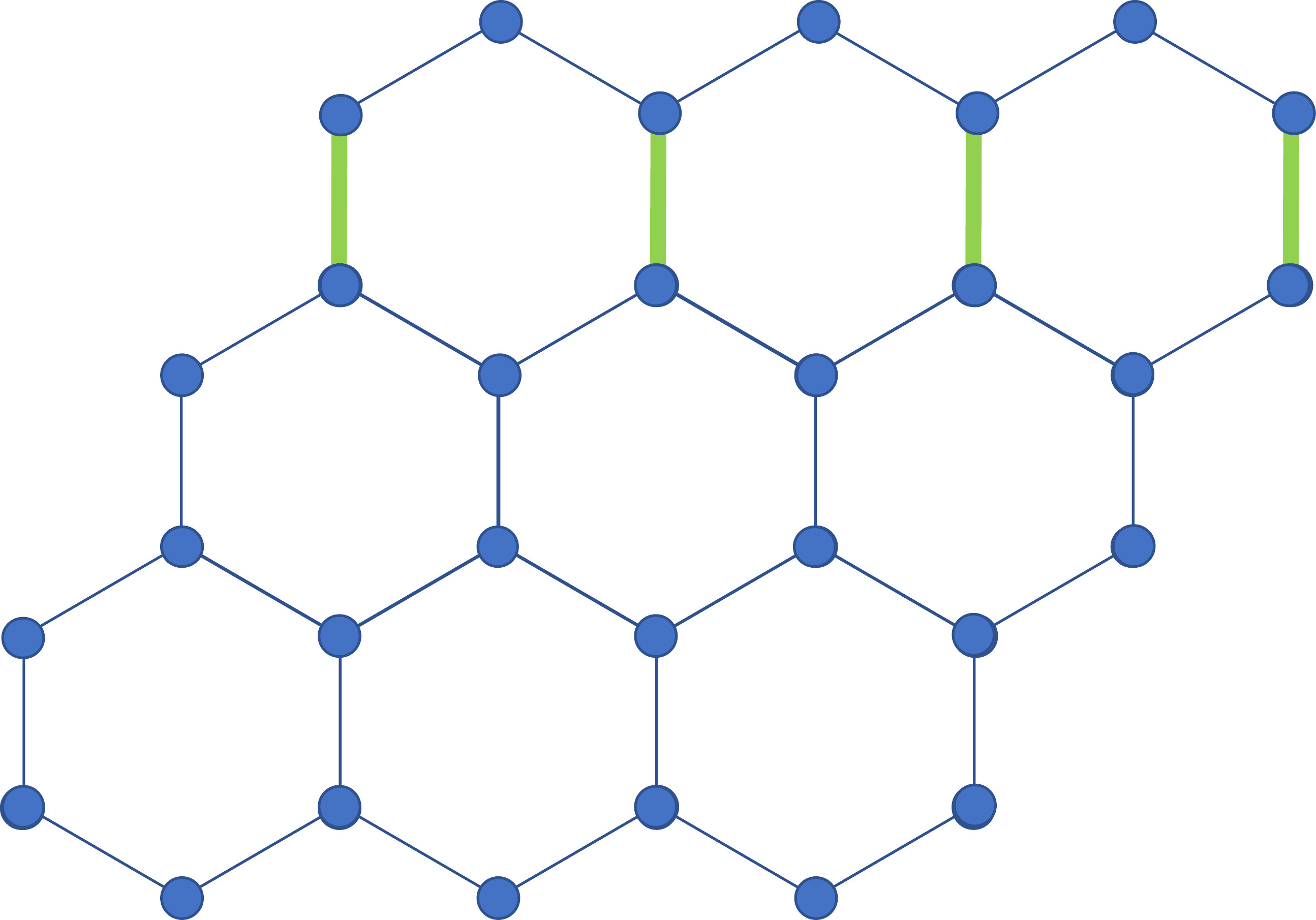}
  \caption{}
  \label{fig:zlinksflipped}
\end{subfigure}
\caption{Simple sets of $y$ (above) and $z$ (below) link flips that preserve the vortex sector while being gauge inequivalent to the trivial link sector.}
\label{fig:ylinksandzlinksflipped}
\end{figure}

\begin{figure}[h]
\begin{subfigure}{.5\textwidth}
  \centering
  % include first image
  \includegraphics[width=.5\linewidth]{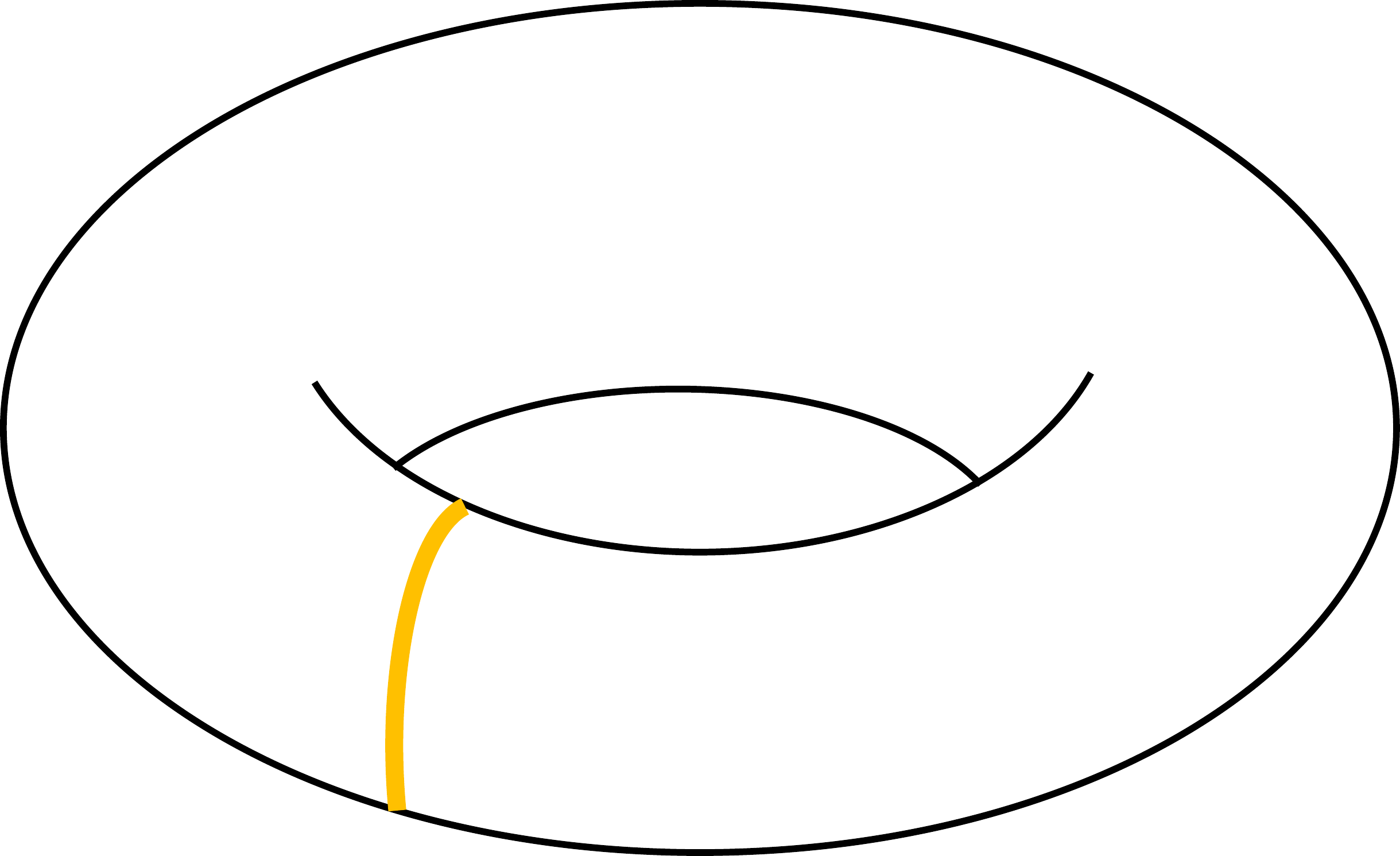}
\caption{}
  \label{fig:ylinksflippeddonut}
\end{subfigure}
\begin{subfigure}{.5\textwidth}
  \centering
  % include second image
  \includegraphics[width=.5\linewidth]{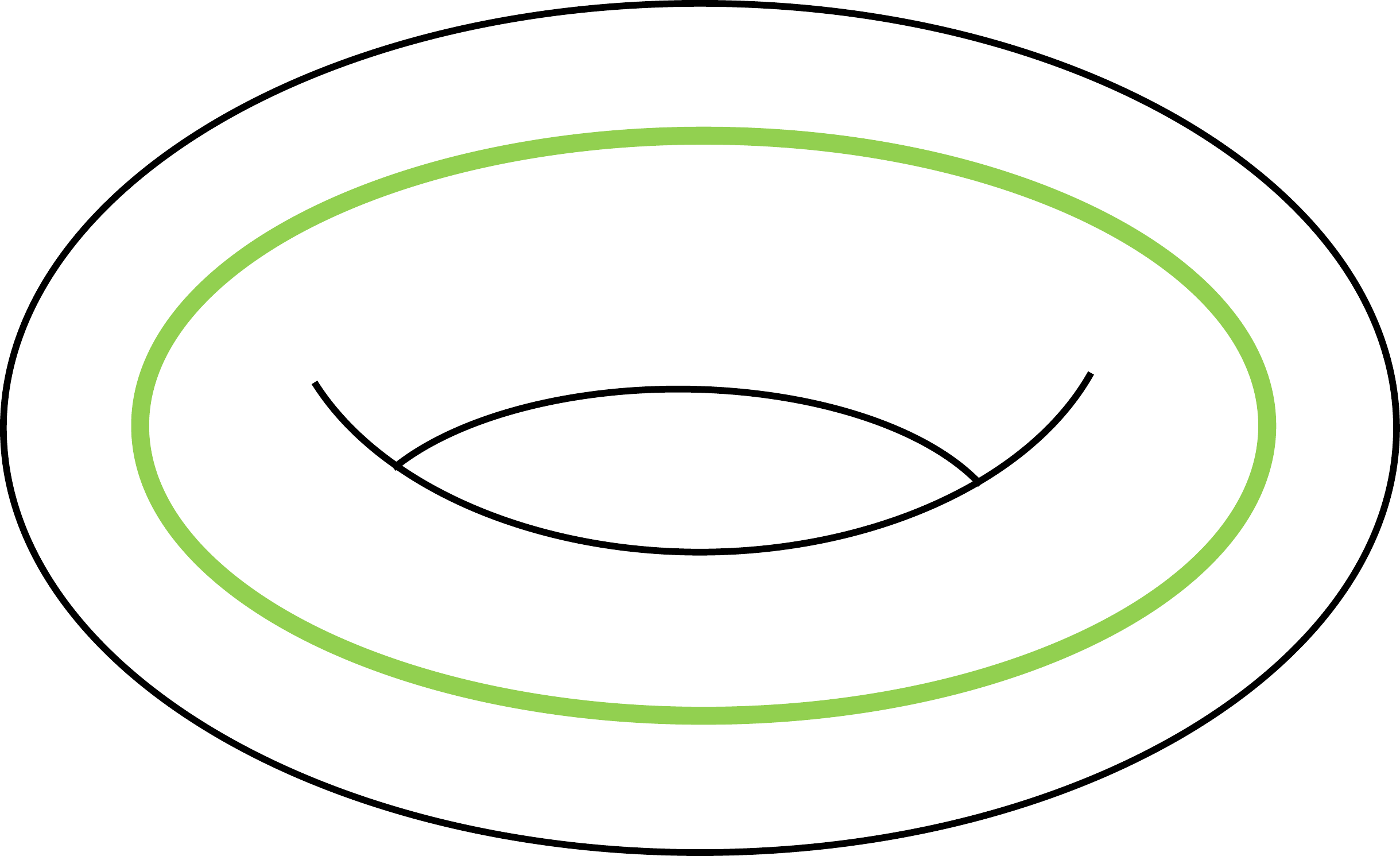}
  \caption{}
  \label{fig:zlinksflippeddonut}
\end{subfigure}
\caption{Topological interpretation of the set of $y$ and $z$ flips that preserve vortex sector but are gauge inequivalent to the trivial link sector which topologically amounts to no loop at all.}
\label{fig:donutylinksandzlinksflipped}
\end{figure}

We now consider the case of a lattice with periodic boundary conditions in both the $a$ and $b$ directions (that is, both vertically and horizontally as on a torus). We consider the no-vortex sector and and any arbitrary link sector associated with it which we call $P_{u_0}$. We have already shown that $P_{u_0}$ is gauge equivalent, that is, equivalent up to conjugation by $\Gamma_k$ operators, with $2^{N-1}$ link sector projectors.

Acting on a non-zero eigenstate of $P_{u_0}$ with any stabilizer operator $D_j$ will change three link sector eigenvalues at a time, due to $D_j$ overlapping with the three types of link that contain spin site $j$, the same is therefore also true of $\Gamma_k$ operators. On the other hand acting on spin sites with a Pauli operator such as $Z_j$, which in the fermionic picture is
\begin{align}
    \tilde{Z}_j = ib^z_j c_j
\end{align}
will flip the sign of only the $z$-link corresponding to that site and no others. In general a Pauli operator $\tilde{\sigma}^\alpha$ will only flip the sign of the $\alpha$-link containing spin site $j$. This is due to the fact that
\begin{align}
    \{ \tilde{\sigma}_j ,  \hat{u}_{jk} \} &= \{ ib^{\sigma}_j c_j ,  ib_j^{\sigma} b_k^{\sigma} \} = 0 \\
    \left[ \tilde{\sigma}_j ,  \hat{u}_{kl} \right] &= \left[ ib^{\sigma}_j c_j ,  ib_k^{\sigma} b_l^{\sigma} \right] = 0, ~~~ j \notin \{k,l\}.
\end{align}

Thus we have that
\begin{align}
    \tilde{Z}_\alpha P_{u_j} = P_{u_k} \tilde{Z}_\alpha
\end{align}
and thus
\begin{align}
    P_{u_j} = \tilde{Z}_\alpha P_{u_k} \tilde{Z}_\alpha
\end{align}
where in general $\omega(u_j) \neq \omega(u_k)$.

Flipping a single link eigenvalue will change the vortex sector as a pair of vortices are introduced on adjacent plaquettes. Taking $\zeta$ to be an arbitrary product of $\tilde{Z}$ Pauli operators, then in order to have
\begin{align}
     P_{u_j} = \zeta P_{u_k} \zeta
\end{align}
such that $\omega(u_j)$ \textit{does} equal $ \omega(u_k)$, we require $\zeta$ to consist of a pair of link-flipping $\tilde{Z}$ operators for each plaquette. 

In order to return to the no-vortex sector, vortices need to be annihilated and so the vortex string must form a closed loop.
Algebraically speaking, while each $\tilde{\sigma}_j$ will commute with all overlapping and non-overlapping $D_j$ operators, as
\begin{align}
    \left[ \tilde{\sigma}_j, D_j \right] &= \left[ ib^\alpha_j c_j , b_j^x b_j^y b_j^z c_j  \right] \\ &= 0
\end{align}
and
\begin{align}
    \left[ \tilde{\sigma}_j, D_k \right] &= \left[ ib^\alpha_j c_j , b_k^x b_k^y b_k^z c_k  \right] \\ &= 0,
\end{align}
such Paulis will not commute with overlapping plaquette operators $W_p$ as for a spin site $j$, plaquette operators will act with Paulis $\tilde{X}$ or $\tilde{Y}$ and thus there is anticommutation.

\begin{figure}[h]
    \centering
    \includegraphics[width=2cm]{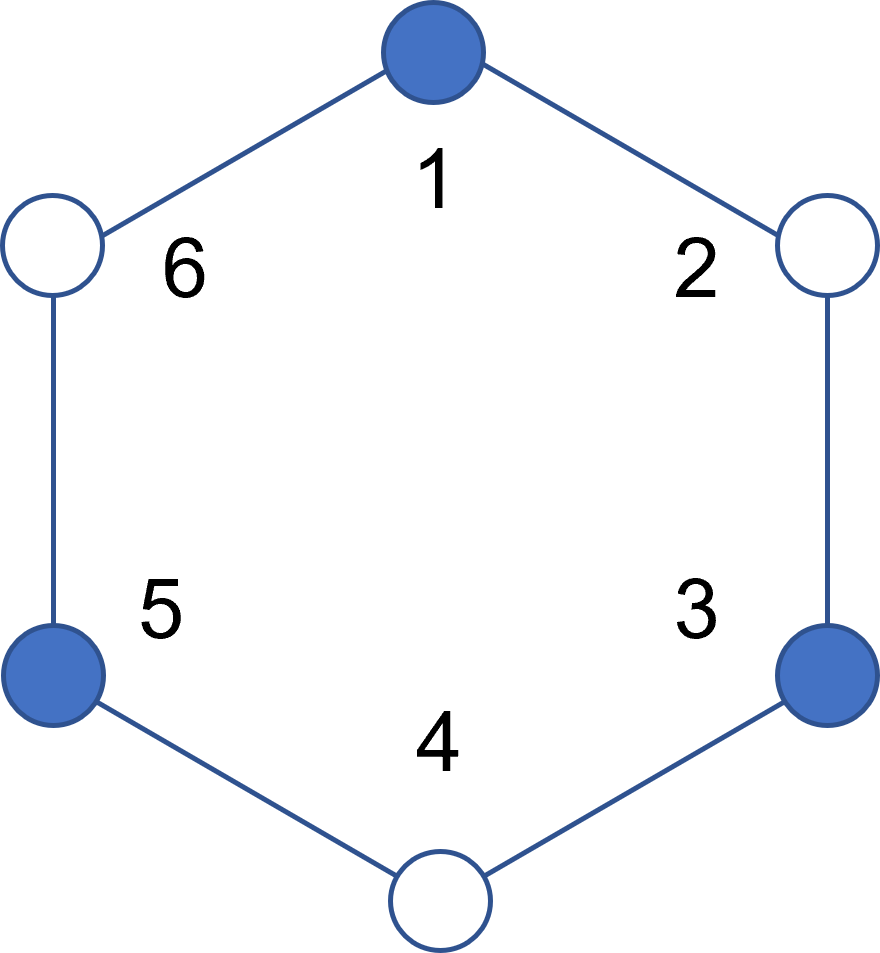}
    \caption{Clearly $\left[ Z_6, W_p\right] = \left[ Z_6, Z_1X_2Y_3Z_4X_5Y_6\right] \neq 0$ but $\left[ Z_6Z_2, W_p\right] = \left[ Z_6Z_2, Z_1X_2Y_3Z_4X_5Y_6\right] = 0$.}
    \label{fig:plaq1to6}
\end{figure}

In order for there to be commutation with all plaquette operators, there needs to be $\tilde{Z_j}$ operators acting on two spin sites per plaquette, as shown in Fig.~\ref{fig:plaq1to6}. The simplest example of a product of $\tilde{Z}_j$ operators that commutes with all link operators is shown explicitly and schematically in Fig.~\ref{fig:zlinksflipped} and Fig.~\ref{fig:zlinksflippeddonut} respectively. 

\begin{figure}[h]
    \centering
    \includegraphics[width=5cm]{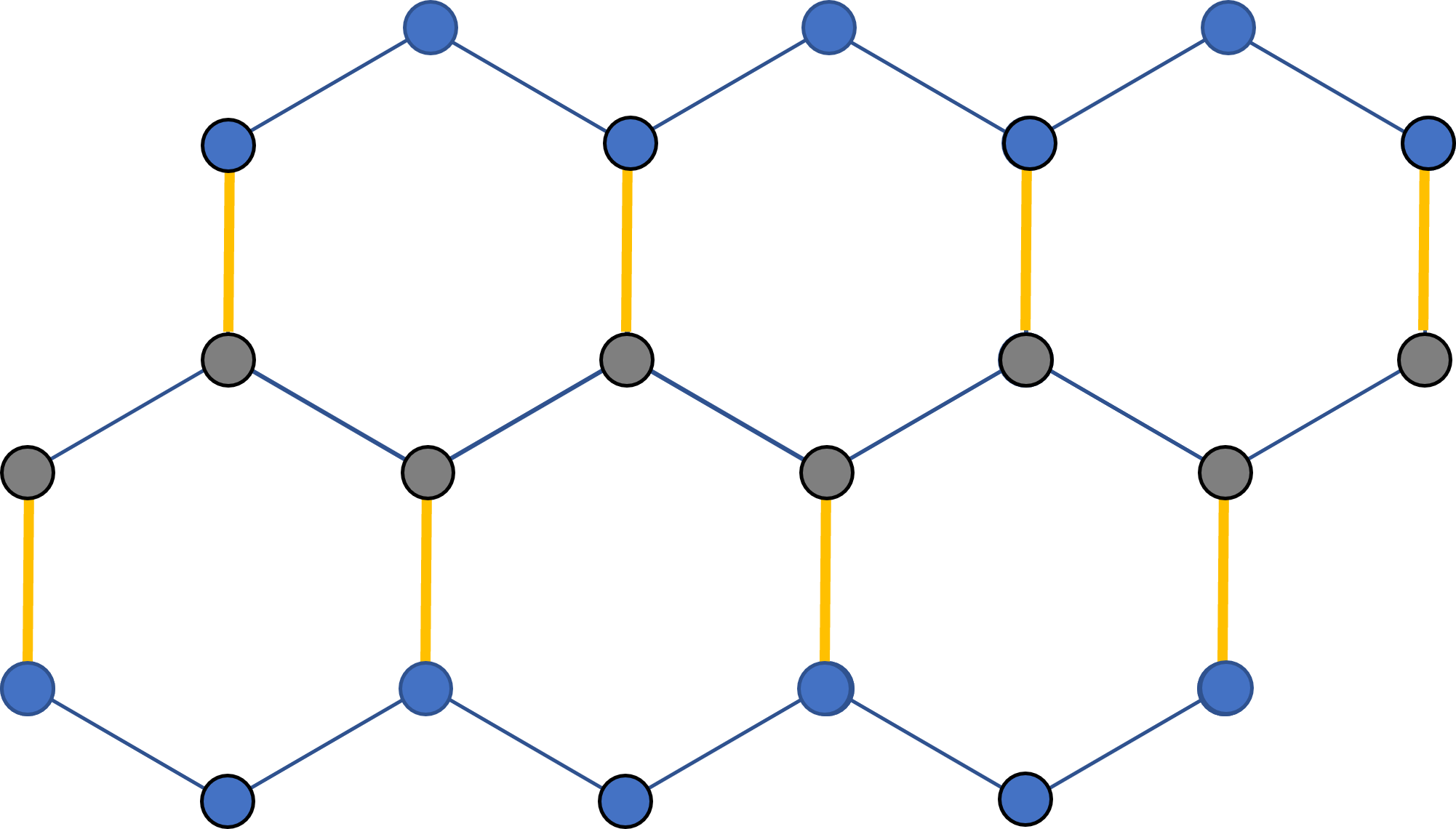}
    \caption{A lattice where the grey spin sites have been acted on with $D_j$ operators and link eigenvalues on yellow links have been flipped. It is not possible to flip only a single line of $z$-links using $\Gamma_k$ operators alone.}
    \label{fig:rows2linksflipped}
\end{figure}

Starting with link sector $P_{u_0}$ and using only $\Gamma_k$ operators, it is not possible to flip the sign of only a single row of $z$-links as in general the action of a $D_j$ operator flips the sign of all three types of link $x,y$ and $z$. In order to keep all $x$ and $y$ links unflipped, products of $D_j$ operators must act on the sites of all spins that correspond to those links, leaving at least a \textit{pair} of rows with flipped eigenvalues on $z$-links, as demonstrated in Fig.~\ref{fig:rows2linksflipped}. We therefore have found another link sector $P_{u_z}$ whose corresponding vortex sector is the no-vortex sector which cannot be reached from $P_{u_0}$ by conjugation with $\Gamma_k$ operators.
Algebraically this can be represented by
\begin{align}
    \nexists~ \Gamma_k ~ \text{such that}~ \Gamma_k P_{u_0} \Gamma_k = P_{u_z}
\end{align}
where
\begin{align}
    P_{u_z} := \zeta P_{u_0} \zeta
\end{align}
and $\zeta$ is a horizontal, topologically closed loop of Pauli $\tilde{Z}$ operators. This is because when commuting $\zeta$ through $P_{u_0}$, only the signs of a row of $z$-link eigenvalues are changed. If we call this row of links $\rho$ then
\begin{align}
    \zeta P_{u_0} &= \zeta \prod_{jk} \frac{1+u^{\alpha}_{jk} \hat{u}^{\alpha}_{jk}}{2} \\
    &= \prod_{\alpha=z, {jk} \in \rho} \frac{1-u^{\alpha}_{jk} \hat{u}^{\alpha}_{jk}}{2}  \prod_{\text{etc.}} \frac{1+u^{\alpha}_{jk} \hat{u}^{\alpha}_{jk}}{2} \zeta \\
    &= P_{u_k} \zeta.
\end{align}

However with $\Gamma_k$ operators, in order to commute them through $P_{u_0}$ and have only $z$-link eigenvalues flipped, this requires a $\Gamma_k$ consisting of $D_j$ operators acting on a row of spin sites in the manner of Fig.~\ref{fig:rows2linksflipped} and thus a minimum of two rows of $z$-link eigenvalues are flipped. Thus for all $\Gamma_k$
\begin{align}
    \Gamma_k P_{u_0} \Gamma_k \neq \zeta P_{u_0} \zeta.
\end{align}

We can now repeat the above process, with the same reasoning, starting with the link sector $P_{u_z}$ rather than $P_{u_0}$, but this time
with Pauli $\tilde{Y}$  operators to flip links as shown in Fig.~\ref{fig:ylinksflipped} and Fig.~\ref{fig:ylinksflippeddonut}. This would have given another link sector unreachable via actions of $\Gamma_k$ operators, just as $P_{u_z}$ was unreachable from $P_{u_0}$. This would have corresponded to flipping link signs vertically rather than horizontally. Pauli $\tilde{X}$ operators can be written, up to a phase, as simply products of $\tilde{Z}$ and $\tilde{Y}$ Paulis, thus no new link sectors can be found by use of the above operations with $\tilde{X}$ Paulis. Therefore for a lattice defined by $a$ plaquettes per row and made up of $b$ rows, if we define 
\begin{align}
    \zeta &= \tilde{Z}_{j_1} ... \tilde{Z}_{j_a} \\
    \chi &= \tilde{Y}_{j_1} ... \tilde{Y}_{j_b}
\end{align}
then the four link sector operators
\begin{align}
 P_{u_0} &, \\
 P_{u_z} & = \zeta P_{u_0} \zeta, \\
 P_{u_y} & = \chi P_{u_0} \chi, \\
 P_{u_x} & = \zeta \chi P_{u_0} \chi \zeta
\end{align}
each define gauge-inequivalent link sectors in a particular vortex sector.

\subsubsection*{Inequivalent sectors: Half-periodic boundary conditions}

For a system with half-periodic boundary conditions, depending on whether periodicity is in the horizontal or vertical direction, only a single row or `column' of flipped signs would be required to find a link sector projector not reachable from $P_{u_0}$ by actions of $\Gamma_k$ operators. There would therefore be two sets of $2^{N-1}$ link sector projectors corresponding to two equivalence classes in each vortex sector. Accounting for all $2^{P}$ vortex sectors there would be a total of
\begin{align}
    2^P (2 \times 2^{N-1}) = 2^{N+P} = 2^L
\end{align}
and so we have accounted for all link sectors.

\section{Proof of Heisenberg-infidelity state-infidelity bound}

This section contains the proof of Eq~\eqref{bound} which relates state infidelity and Heisenberg infidelity. State infidelity is defined between two states with differing dynamics $U_j~ (j=1,2)$ applied to an initial spin state $\ket{\Psi_I}$. These dynamics are induced by spin Hamiltonians $H_{s,j}$ whose counterparts in the quadratic Majorana fermionic picture  $\tilde{H}_{\text{ferm},j} = \alpha^\dagger M_j \alpha$ (see Section.~\ref{Majoranadiag}) define orthogonal matrices $\mathcal{O}_j$ generated by the matrices $M_j$ in a relation which satisfies the differential equation $\dot{\mathcal{O}}_j = iM_j \mathcal{O}_j$.

The relation to be proven reads
\begin{align}
    \mathcal{I}_H (\mathcal{O}_1, \mathcal{O}_2) \geq \frac{1}{4d^3}\left(1-\sqrt{\mathcal{F}_s (\Psi_1,\Psi_2})\right) 
\end{align}
with the state fidelity
\begin{align}
 \mathcal{F}_s(\Psi_1,\Psi_2)\ = \left| \ovl{\Psi_1}{\Psi_2} \right|^2\ = \left| \bra{\Psi_I} U_2^\dagger U_1 \ket{\Psi_I} \right|^2,
 \label{statefidelity}
\end{align}
and the  Heisenberg infidelity $\mathcal{I}_H=1-\mathcal{F}_H$ with
\begin{align}
    \mathcal{F}_H(\mathcal{O}_1, \mathcal{O}_2)  = \frac{1}{d} \left| \Tr(\mathcal{O}_1^\dagger  \mathcal{O}_2 )\right|\ ,
\end{align}
and the matrix dimension $d$.

The full proof consists of Eq.~\eqref{ineq1} and the series of inequalities Eq.~\eqref{ineq2} to \eqref{ineq5}
\begin{align}
    \sqrt{2d} \sqrt{\mathcal{I}_H} & =\| \mathcal{O}_1 - \mathcal{O}_2 \|_F\label{ineq1}\\
    &\geq \frac{1}{2d}   \| \Phi_{U_1} - \Phi_{U_2} \|_\lozenge \label{ineq2}\\
    &\geq \frac{1}{2d} \min_\varphi \| U_1 - e^{i\varphi} U_2 \|_\text{op}\label{ineq3}\\
    &\geq \frac{1}{\sqrt{2}d} \sqrt{1-\sqrt{\mathcal{F}_s}}\label{ineq5}
\end{align}
that will be discussed separately in the following subsections.

Eq.~\eqref{ineq1} is expressed in terms of the Frobenius norm %$\| \circ \|_F$ is
\begin{align}
\| A \|_F :=\sqrt{\Tr(A^\dagger A)}
\end{align}
for any operator $A$. 
Eq.~\eqref{ineq2} is expressed in terms of the operator norm
\begin{align}
\|A\|_\text{op} &:= \sup\left\{\frac{\|Ax\|}{\|x\|} : x\in V^d \text{ with }x\ne 0\right\}.
\label{eq:operatornorm}
\end{align} 
The diamond norm for a quantum channel $\Phi$ in Eq.~\eqref{ineq3} is given by
\begin{align}
    \| \Phi \|_\lozenge := \max_\rho \| (\Phi \otimes \mathds{1} ) \rho \|_1
\end{align}
where $\| \circ \|_1=\Tr\sqrt{A^\dagger A}$ is the trace norm~\cite{Benenti2010ComputingRepresentation}, and the maximization is taken over all density matrices in a space of dimension corresponding to the size of the quantum channel.

\subsection{ \eqref{ineq1}}

The Frobenius norm of the difference between two orthogonal operators $\mathcal{O}_1$ and $\mathcal{O}_2$ reduces to
\begin{align}
\| \mathcal{O}_1-\mathcal{O}_2 \|_F &= \sqrt{2d-\Tr(\mathcal{O}_1^\mathrm{T} \mathcal{O}_2)-\Tr(\mathcal{O}_2^\mathrm{T} \mathcal{O}_1)}\ \\
&= \sqrt{2d-2 \Re{ \Tr(\mathcal{O}_1^\mathrm{T} \mathcal{O}_2)}},
\end{align}
where $d$ is the dimension of $\mathcal{O}_1$ and $\mathcal{O}_2$.
It thus depends on the real part of $\Tr(\mathcal{O}_1^\mathrm{T} \mathcal{O}_2)$ and not on its absolute value as it is the case for $\mathcal{F}_H$.
As we will show in the following, however, in the present case, the object $\Tr(\mathcal{O}_1^\mathrm{T} \mathcal{O}_2)$ is real and positive, so that Eq.~\eqref{ineq1} is indeed satisfied.

\subsubsection{Proof that $\Tr(\mathcal{O}_1^\mathrm{T} \mathcal{O}_2)$ is real}

The orthogonal matrices $\mathcal{O}_1$ and $\mathcal{O}_2$ 
satisfy the differential equation $\dot{\mathcal{O}}_j=iM_j\mathcal{O}_j$ with generally time-dependent generators $iM_j$.
Since the $M_j$ are purely imaginary, the generators $iM_j$ are purely real.
Together with the initial condition $\mathcal{O}_j(0)={\mathds{1}}$, such that $\mathcal{O}_j(0)$ is real, this implies that $\mathcal{O}_j(t)$ for $j=1,2$ is real for all times.
Consequently the overlap $\Tr(\mathcal{O}_1^\mathrm{T} \mathcal{O}_2)$ is also real.

\subsubsection{Proof that $\Tr(\mathcal{O}_1^\mathrm{T} \mathcal{O}_2)$ is non-negative}

Since Majorana fermions move between only one of two pairs of fermionic sites per spin site,
the full space that $\mathcal{O}_1$ and $\mathcal{O}_2$ act on, can be divided into two subspaces $H_x$ and $H_y$ of equal dimension $d/2$.
Both $\mathcal{O}_1$ and $\mathcal{O}_2$ are given as a direct sum of the identity ${\mathds{1}}$ in $H_x$ and orthogonal operators $\tilde{\mathcal{O}}$ in $H_y$.
The complete trace is thus given as the sum of the two traces $\Tr_x{\mathds{1}}$ and $\Tr_y\tilde{\mathcal{O}}_{1}^\mathrm{T}\tilde{\mathcal{O}_2}$.
The latter trace can also be expressed as the sum over the eigenvalues $\lambda_j$ of $\tilde{\mathcal{O}}_{1}^\mathrm{T}\tilde{\mathcal{O}_2}$.
This results in the relation
\begin{align}
\Tr(\mathcal{O}_1 ^\mathrm{T} \mathcal{O}_2)=
\Tr_x{\mathds{1}}+\Tr_y \tilde{\mathcal{O}}_{\text1}^\mathrm{T}\tilde{\mathcal{O}_2}=
\frac{d}{2}+\sum_{j=1}^{\frac{d}{2}}\lambda_j\ .
\end{align}
Since the trace $\Tr(\mathcal{O}_1 ^\mathrm{T} \mathcal{O}_2)$ is purely real, the imaginary parts $\Im\lambda_j$ add up to zero.
Because all the eigenvalues $\lambda_j$ are phase factors, {\it i.e.} $\lambda_j=\exp(i\varphi_j)$ with $\varphi_j$ real, the real parts of the $\lambda_j$ satisfy the inequality $\Re\lambda_j\ge -1$,
so that the relation
\begin{align}
\Tr(\mathcal{O}_1 ^\mathrm{T} \mathcal{O}_2)=\frac{d}{2}+\sum_{j=1}^{\frac{d}{2}}\lambda_j\ge 0
\end{align}
is indeed given.

\subsection{from \eqref{ineq1} to \eqref{ineq2} }

The Frobenius and operator matrix norms satisfy the inequality $\|A\|_F \geq \|A\|_\text{op}$~\cite{Golub1996MatrixComputations}.
For the present case, this implies the relation
\be
\| \mathcal{O}_1 - \mathcal{O}_2 \|_F \ge \| \mathcal{O}_1 - \mathcal{O}_2 \|_\text{op}\ .
\ee
Together with the relation
\begin{align}
    \| \mathcal{O}_1 - \mathcal{O}_2 \|_\text{op} \geq \frac{1}{2d}   \| \Phi_{U_1} - \Phi_{U_2} \|_\lozenge \ ,
\end{align}
that is proven in Eq.~(171) in~\cite{Oszmaniec2022FermionStates} this yields the desired inequality.

\subsection{from \eqref{ineq2} to \eqref{ineq3} }

The required inequality
\begin{align}
     \| \Phi_{U_1} - \Phi_{U_2} \|_\lozenge &\geq  \min_\varphi \| U_1 - e^{i\varphi}U_2 \|_\text{op}
\end{align}
is proven in Eq.~(2.1) in~\cite{Oszmaniec2022Epsilon-NetsCircuits}

\subsection{from \eqref{ineq3} to \eqref{ineq5} }

Following the definition of the operator norm (Eq.~\eqref{eq:operatornorm}), the operator norm in Eq.~\eqref{ineq3} satisfies the inequality
\begin{align}
\| U_1 - e^{i\varphi}U_2 \|_\text{op}\geq  \| (U_1 - e^{i\varphi} U_2) \ket{\Psi} \|_2
\end{align}
for any state vector $\ket{\Psi}$ in the spin picture, and, as such, in particular for the initial state $\ket{\Psi_I}$ of the dynamics.
That is, the inequality
\begin{align}
\| U_1 - e^{i\varphi}U_2 \|_\text{op} &\geq  \| \ket{\Psi_1} - e^{i\varphi} \ket{\Psi_2}  \|_2\\
&= \sqrt{(\bra{\Psi_1} - e^{-i\varphi} \bra{\Psi_2}) (\ket{\Psi_1} - e^{i\varphi} \ket{\Psi_2}}) \\
&= \sqrt{2 - e^{i\varphi} \braket{\Psi_1}{\Psi_2} - e^{-i\varphi} \braket{\Psi_2}{\Psi_1}}\\
&= \sqrt{2 - 2 | \ovl{\Psi_1}{\Psi_2}|\cos(\varphi + \theta)}
\label{twominussumofoverlaps}
\end{align}
holds, where the overlap of $\ket{\Psi_1}$ and $\ket{\Psi_2}$ can be defined in terms of state infidelity and a phase
\begin{align}
    \ovl{\Psi_1}{\Psi_2} = | \ovl{\Psi_1}{\Psi_2}|e^{i\theta} = \sqrt{\mathcal{F}_s} e^{i\theta}.
\end{align}
When Eq.~\eqref{twominussumofoverlaps} is minimized over all phases $\varphi$ this gives the required result
\begin{align}
   \min_\varphi \| U_1 - e^{i\varphi}U_2 \|_\text{op} &\geq \sqrt{2}\sqrt{1- \sqrt{\mathcal{F}_s (\Psi_1,\Psi_2})}.
\end{align}

\end{appendix}

\bibliography{references.bib}

\end{document}